\documentclass[preprint,12pt]{elsarticle}

\usepackage{amssymb}

\usepackage{amsmath}

\usepackage[psdextra]{hyperref}

\usepackage{subcaption}

\usepackage{siunitx}

\journal{Journal of Sound and Vibration}

\begin{document}

\begin{frontmatter}



\title{Uncertainty Analysis of Limit Cycle Oscillations in Nonlinear Dynamical Systems with the Fourier Generalized Polynomial Chaos Expansion}

\author[inst1]{Lars de Jong\corref{cor1}}
\ead{l.de-jong@tu-braunschweig.de}
\cortext[cor1]{Corresponding author}

\affiliation[inst1]{organization={Institute for Acoustics and Dynamics, TU Braunschweig},
            addressline={Langer Kamp 19}, 
            city={Braunschweig},
            postcode={38106}, 
            country={Germany}}

\author[inst1]{Paula Clasen}
\author[inst1]{Michael Müller}
\author[inst1]{Ulrich Römer}

\begin{abstract}

In engineering, simulations play a vital role in predicting the behavior of a nonlinear dynamical system. In order to enhance the reliability of predictions, it is essential to incorporate the inherent uncertainties that are present in all real-world systems. Consequently, stochastic predictions are of significant importance, particularly during design or reliability analysis. In this work, we concentrate on the stochastic prediction of limit cycle oscillations, which typically occur in nonlinear dynamical systems and are of great technical importance. 
To address uncertainties in the limit cycle oscillations, we rely on the recently proposed Fourier generalized Polynomial Chaos expansion (FgPC), which combines Fourier analysis with spectral stochastic methods. 
In this paper, we demonstrate that valuable insights into the dynamics and their variability can be gained with a FgPC analysis, considering different benchmarks. These are the well-known forced Duffing oscillator and a more complex model from cell biology in which highly non-linear electrophysiological processes are closely linked to diffusive processes.
With our spectral method, we are able to predict complicated marginal distributions of the limit cycle oscillations and, additionally, for self-excited systems, the uncertainty in the base frequency. Finally we study the sparsity of the FgPC coefficients as a basis for adaptive approximation. 

\end{abstract}

\begin{highlights}
\item Combination of Fourier analysis with Polynomial Chaos expansion 
\item Efficient uncertainty quantification of base frequency in self-excited systems
\item Sparsity in the Fourier-Polynomial Chaos expansion is demonstrated for given examples
\end{highlights}


\end{frontmatter}


\section{Introduction}
\label{sec:Introduction}

Stochastic predictions play a pivotal role for numerous systems, as these systems exhibit aleatory uncertainty in reality. Depending on the uncertainty, this can significantly affect the dynamic behavior of the system. This especially holds true for nonlinear systems. Here, small changes in deterministic values of the system parameter can have a significant impact on the behavior \cite{parkerPracticalNumericalAlgorithms1989a}.

Therefore, uncertainty must be accounted for in different aspects of a system's evaluation, e.g., in a design process, in the reliability analysis of a product, or in the individual treatment of a patient in the field of medicine. For the given examples, it is crucial to quantify uncertainties in nonlinear dynamical systems. For the design process, this can mean to account for varying material properties. With regard to reliability, this could be the probability of exceeding a maximum amplitude for a given number of cycles - and in individual treatment, this could concern the effect of a medicine on different human body conditions.

Real world experiments contain uncertainties as well. If the number of experiments were to be decreased, simulations would usually replace some or most of these experiments. Most real-world experiments can be described using nonlinear physical relations. To account for a wide range of real-world behavior, uncertainties need to be added to the simulation model.

Most systems operate in a region which can be considered as steady state. For nonlinear systems, this usually means they end in stable equilibrium points or limit cycle oscillations (LCO). We focus on the latter one. The process to reach this kind of state in simulations can take a long time, especially if the system requires significant computational resources. As a means to avoid this long process and perform a direct analysis of the steady state, the Harmonic Balance method (HB) is well known and established. It originated from the evaluation of electric circuits \cite{gilmoreNonlinearCircuitAnalysis1991a,gilmoreNonlinearCircuitAnalysis1991b,sandersLimitCyclesDescribing1993} but was also well adapted for problems in the field of computational fluid dynamics \cite{hallComputationUnsteadyNonlinear2002} and mechanical problems \cite{krackMethodNonlinearModal2013a}. Another advantage of the HB compared to the time integration methods is its ability to find unstable solutions within the system. 

As mentioned before, uncertainties can significantly impact the behavior of nonlinear systems. Attempts to quantify uncertainties in nonlinear systems are not necessarily new \cite{beranUncertaintyQuantificationLimitcycle2006, lemeitourPredictionStochasticLimit2010}, but have experienced growing interest in recent years, both in Mathematics \cite{bredenComputingInvariantSets2020,kuehnUncertaintyQuantificationAnalysis2024} and in Engineering 
\cite{trinhNonlinearDynamicBehaviour2020,thomasNonIntrusivePolynomialChaos2022}.

The Monte Carlo (MC) method is still the most popular method for quantifying uncertainty. However, it requires that the system is evaluated for a large set of samples of the uncertain parameter in order to estimate the statistics of the limit cycle. Depending on the system, this can be quite time-consuming. Efficient alternatives are, among others, quasi Monte Carlo \cite{dick2013high}, multifidelity approaches \cite{peherstorfer2018survey} or surrogate modeling \cite{gramacy2020surrogates}. Here, we are interested in surrogate modeling in particular, as a means to drastically reduce the simulation time. In uncertainty quantification, one well-established method for surrogate modeling is the generalized Polynomial Chaos expansion (gPC) \cite{xiu2002wiener,ghanem2003stochastic}. This spectral method has proven to be efficient in a wide variety of applications and may be employed in either an intrusive or non-intrusive way \cite{xiu2010numerical}.

It has been noted several times that gPC can be seen as a generalized Fourier expansion, where the polynomials of the underlying random variables are orthogonal, see \cite{nobile2009analysis} for instance. As both HB and gPC employ an orthogonal basis, it appears natural to combine the HB with the intrusive gPC to form the Fourier generalized Polynomial Chaos expansion (FgPC). However, this direct combination is still a relatively novel concept. 

Before that, HB and gPC were partially used together in elaborating uncertainties in nonlinear systems. For example \citet{beranUncertaintyQuantificationLimitcycle2006} used the gPC expansion to calculate statistics of the LCO via the cyclic specified amplitude method. \citet{lemeitourPredictionStochasticLimit2010,chassaingStochasticNonlinearAeroelastic2012} calculated the response surface of a nonlinear system using a gPC expansion. \citet{hayesPredictionLimitCycle2015} used the HB method to generate samples for the non-intrusive gPC expansion in order to describe the uncertainty of the systems solution. 

To the best of our knowledge, the first paper to describe a direct combination of HB and gPC in an intrusive approach was published by 
\citet{sarrouyStochasticStudyNonlinear2013a}, introducing the description of the Fourier coefficients of the HB method through a gPC expansion and describing the stochastic behavior of a nonlinear self-excited system.
In \cite{lamrhariFREQUENCYANALYSISNONLINEAR2017}, the authors employed this approach to analyze the uncertainty of parameters and their effects on the invariants of the solution manifold in dynamical systems. 
\citet{trinhNonlinearDynamicBehaviour2020} describe a clutch system with uncertainties and FgPC, where they account for some uncertain system parameters but only one harmonic for each degree of freedom.
\citet{bredenComputingInvariantSets2020} employed the FgPC methodology to delineate the uncertainty associated with periodic orbits of the Lorenz system. \citet{thomasNonIntrusivePolynomialChaos2022} describe the influence of uncertainties on aeroelastic LCO using the FgPC method with the non-intrusive approach for one harmonic.

So far, FgPC was applied both to academic examples and to fairly simple mechanical systems, as stated above. When functions with low regularity are present, a high number of harmonics is needed to capture the system behavior. Also, when complicated marginal distributions are present, the polynomial degree needs to be increased. Therefore, it is necessary to demonstrate the application and efficiency of this method for complex problems.

The scope of this paper is to extend the work done by \citet{bredenComputingInvariantSets2020}, showing that complex systems can be efficiently described using the FgPC method. We will also demonstrate that there is sparsity in the FgPC coefficients.
The FgPC method focuses on LCO, where it can reach its full potential. On the one hand, with random variables in the system parameters, FgPC returns a random process describing the uncertainties in the investigated systems. On the other hand, for self-excited systems, it additionally returns the frequency as a random variable. The random process captures the marginal distributions of the system's solution for any given time, which is not possible with linear methods.

Nonlinear systems can have multiple solutions for the same set of system parameters. For each solution, we are interested in a stochastic evaluation. Therefore, we will combine the FgPC method with a deflation technique. The deflation technique prevents the root-finding algorithm within the FgPC method from finding the same solution again - ensuring that multiple, or ideally all, solutions can be found.

Additionally, we are providing a suitable code framework for the efficient combination of HB and gPC, which has not been done before, see \cite{de_jong_2024_13359870}.

We apply our FgPC method to two examples. The first example is the well-known Duffing oscillator. The second example is a problem in cell biology. In this case, highly non-linear ion currents in beta cells lead to complex dynamic processes through channels in conjunction with chemical reactions. These also include a multi-layered bifurcation problem, which determines the occurrence of action potentials. The resulting oscillations in the electrical membrane potential are characterized by corresponding limit cycles. The associated uncertainties are addressed in this context. This example also illustrates the wide range of applications of the FgPC.

The structure of this article is as follows. First we will briefly recall both the HB and the gPC method in Section \ref{sec:FgPC}, followed by a combination of both. At the end of Section \ref{sec:FgPC} the deflation technique is explained briefly. Section \ref{sec:duffing} first introduces all important implementation aspects of the method and covers the application of the FgPC method to the Duffing oscillator and the cell biology example. Finally, some conclusions are drawn. 

\section{Fourier Generalized Polynomial Chaos Expansion}
\label{sec:FgPC}

We consider the nonlinear dynamical system
\begin{equation}
    \dot{\boldsymbol{x}} = \boldsymbol{f}(\boldsymbol{x},t,\boldsymbol{\theta}), \ \boldsymbol{x}(0,\boldsymbol{\theta}) = \boldsymbol{x}_0,
    \label{eqn:systemEquation}
\end{equation}
with solution $\boldsymbol{x}(t,\boldsymbol{\theta}) \in \mathbb{R}^{n_{\text{d}}}$, $t \in (0,T_{\text{s}}]$, $\boldsymbol{x}: (0,T_{\text{s}}] \times \mathbb{R}^p \to \mathbb{R}^{n_{\text{d}}}$ and a parameter vector $\boldsymbol{\theta} \in \mathcal{P} \subset\mathbb{R}^p$. Bold symbols represent either a matrix or a vector.

\subsection{Harmonic Balance}
\label{sec:harmonicbalance}

In this section, we fix the parameter vector $\boldsymbol{\theta}$ and omit it to simplify the notation. The effect of stochastic parameter variation will be considered in section \ref{sec:gPC}. 

We assume, that the system has a solution in a quasi-stationary state, which characterizes the periodic solution. Hence, it can be approximated by a complex Fourier series
\begin{equation}
    \boldsymbol{x}(t)= \sum_{k =-\infty}^{\infty} \hat{\boldsymbol{x}}_k e^{i k \omega t}  \approx \boldsymbol{x}_H(t) = \sum_{k =-H}^{H} \hat{\boldsymbol{x}}_k e^{i k \omega t} 
    \label{eqn:fourierSeries}
\end{equation}
truncated after the $H$-th harmonic. Depending on the system, the base frequency $\omega$ is either the excitation frequency or an additional unknown. The latter is the case if the system is self-excited. A Galerkin approach is used to obtain the Fourier coefficients $\hat{\boldsymbol{x}}_k$. Inserting $\boldsymbol{x}_H$ into \eqref{eqn:systemEquation}, we can form the residual $\boldsymbol{r}_{\text{HB}}(\dot{\boldsymbol{x}}_H, \boldsymbol{x}_H,t) = \dot{\boldsymbol{x}}_H - \boldsymbol{f}(\boldsymbol{x}_H,t)$. For complex-valued square-integrable functions $u,v \in L^2([0,T])$, we introduce the inner product specified over one period $T$ of the system
\begin{equation}
    \langle u,v \rangle_{L^2([0,T])} = \frac{1}{T} \int_0^T u(t) \bar{v}(t) \ dt,
\end{equation} 
where $\bar{v}$ denotes complex conjugation. Interpreting $\Psi_l(t) = e^{i l \omega t}$ as basis functions, the Galerkin approach is obtained by equating the inner product between the residual and $\Psi_l$ to zero, i.e., 
\begin{equation}
    \langle r_{\text{HB},c}(\dot{\boldsymbol{x}}_H, \boldsymbol{x}_H,\cdot),\Psi_l \rangle_{L^2([0,T])} = \frac{1}{T} \int_0^T r_{\text{HB},c}(\dot{\boldsymbol{x}}_H, \boldsymbol{x}_H,t) e^{-i l \omega t} dt = 0, 
    \label{eqn:innerproductHB}
\end{equation}
for $l=-H,\ldots,H$. Index $c$ stands for the $c$-th dimension of the system \eqref{eqn:systemEquation} and has the range  $c = 1,\ldots,n_{\text{d}}$ . This results in $n_{\text{d}} (2H+1)$ equations. For a comprehensive treatment of HB, please refer to \cite{krack2019}.

Using a trapezoidal integration or quadrature rule for \eqref{eqn:innerproductHB} results in 
\begin{equation}
    \frac{1}{T} \int_0^T r_{\text{HB},c}(\dot{\boldsymbol{x}}_H, \boldsymbol{x}_H,t) e^{-i l \omega t} dt \approx \frac{1}{N_{t}} \sum_{j=0}^{N_{t}-1} r_{\text{HB},c}(\dot{\boldsymbol{x}}_H(t_j), \boldsymbol{x}_H(t_j), t_j) e^{-i l \frac{2\pi}{N_{t}} j} = 0 \text{,}
    \label{eqn:quadFourierCoeff}
\end{equation}
for $l=-H,\ldots,H$, where $t_j = j 2 \pi /N_{t}$. Here, $N_{t}$ denotes the total number of equidistant points in time within one period $T$.
With the $2H+1$ equations, \eqref{eqn:quadFourierCoeff} can be rewritten in matrix form 
\begin{equation}
\begin{aligned}
    \underbrace{ \frac{1}{N_{t}}
    \begin{bmatrix}
        e^{-i (-H) \frac{2\pi}{N_{t}} 0} & \cdots & e^{-i (-H) \frac{2\pi}{N_{t}} (N_{t}-1)} \vphantom{g(\dot{x}_H(t_0), x_H(t_0), t_0)}\\
        \vdots & & \vdots \\
        e^{-i H \frac{2\pi}{N_{t}} 0} & \cdots & e^{-i H \frac{2\pi}{N_{t}} (N_{t}-1)} \vphantom{g(\dot{x}_H(t_0), x_H(t_0), t_0)}
    \end{bmatrix} \otimes \boldsymbol{I}_{n_{\text{d}}}
    }_{\boldsymbol{E}_{HN_{t}}^{*}} &\\
    \underbrace{
    \begin{bmatrix}
        \vphantom{e^{-i \frac{2\pi}{N_{t}}}} \boldsymbol{r}_{\text{HB}}(\dot{\boldsymbol{x}}_H(t_0), \boldsymbol{x}_H(t_0), t_0) \\
        \vdots \\
        \vphantom{e^{-i \frac{2\pi}{N_{t}}}} \boldsymbol{r}_{\text{HB}}(\dot{\boldsymbol{x}}_H(t_{N_{t}-1}), \boldsymbol{x}_H(t_{N_{t}-1}), t_{N_{t}-1})
    \end{bmatrix}
    }_{\Tilde{\boldsymbol{r}}_{\text{HB}}} &= \boldsymbol{0},
    \label{eqn:MatquadFourierCoeff}
\end{aligned}
\end{equation}
where $\boldsymbol{E}_{HN_{t}}^{*}$ represents in fact the fast Fourier transformation. Also, $\otimes$ is the Kronecker product and $\boldsymbol{I}_{n_{\text{d}}}$ is an identity matrix of size $n_{\text{d}}$. Any symbol with $\Tilde{}$ indicates that it is evaluated at the quadrature points. 

With this representation the residual needs to be evaluated at all time steps $N_{t}$. Therefore, the position $\boldsymbol{x}$ and velocity $\dot{\boldsymbol{x}}$ will be evaluated at these time steps in the time domain as well. This is done by 
\begin{align}
    \Tilde{\boldsymbol{x}}_H &= \underbrace{
    \begin{bmatrix}
        e^{i (-H) \frac{2\pi}{N_{t}} 0} & \cdots & e^{i H \frac{2\pi}{N_{t}} 0} \\
        \vdots & & \vdots \\
        e^{i (-H) \frac{2\pi}{N_{t}}  (N_{t}-1)} & \cdots & e^{i H \frac{2\pi}{N_{t}} (N_{t}-1)}
    \end{bmatrix} \otimes \boldsymbol{I}_{n_{\text{d}}} }_{\boldsymbol{E}_{N_{t}H}} 
     \underbrace{\begin{bmatrix}
        \hat{\boldsymbol{x}}_{-H} \\
        \vdots \\
        \hat{\boldsymbol{x}}_{H}
    \end{bmatrix} }_{\hat{\boldsymbol{x}}} \label{eqn:discretePos}\\
    \dot{\Tilde{\boldsymbol{x}}}_H &= \boldsymbol{E}_{N_{t}H} \omega (\underbrace{\text{diag}(-iH, \cdots, H) \otimes \boldsymbol{I}_{n_{\text{d}}}}_{\boldsymbol{H}} \hat{\boldsymbol{x}} ),
    \label{eqn:discreteVel}
\end{align}
where $\boldsymbol{E}_{N_{t}H}$ can also be seen as the inverse fast Fourier transformation. 
In short, the determining equations for the Fourier coefficients are
\begin{equation}
    \boldsymbol{0} = \boldsymbol{E}_{HN_{t}}^{*} \Tilde{\boldsymbol{r}}_{\text{HB}}(\boldsymbol{E}_{N_{t}H} \omega \boldsymbol{H} \hat{\boldsymbol{x}}, \boldsymbol{E}_{N_{t}H} \hat{\boldsymbol{x}}, \tilde{\boldsymbol{t}}).
    \label{eqn:HBdetEq}
\end{equation}
The roots of \eqref{eqn:HBdetEq} are the corresponding Fourier coefficients $\hat{\boldsymbol{x}}_k$. 

The derived method is called Alternating Frequency Time Harmonic Balance Method (AFTHB), which based on the original idea from \citet{cameronAlternatingFrequencyTime1989}.

\subsection{Generalized Polynomial Chaos expansion}
\label{sec:gPC}

We now model the parameter $\boldsymbol{\theta}$ as a random vector to account for uncertainties in \eqref{eqn:systemEquation}, which also result in an uncertain solution of the system. The random vector $\boldsymbol{\theta}$ takes values in the parameter domain $\mathcal{P} \subset \mathbb{R}^p$, and we assume that the joint probability density function $\pi_{\boldsymbol{\theta}}$ is known, e.g., through parameter estimation with noisy data. The general goal is then to compute moments or other statistical descriptions of the solution random field $\boldsymbol{x}(t,\boldsymbol{\theta})$. Here, we use the gPC expansion with polynomials $\Phi_m(\boldsymbol{\theta})$ chosen according to the Askey scheme \cite{xiuWienerAskeyPolynomial2002}. This ensures that the polynomial basis and, hence, the surrogate is adapted to the density $\pi_{\boldsymbol{\theta}}$. The polynomials are orthonormal in the sense that 
\begin{equation}
    \mathbb{E}[\Phi_m \Phi_n] = \delta_{mn}, \quad \delta_{mn}=0, m\neq n, \text{ and } \delta_{mn}=1, m = n,
\end{equation}
and the coefficients $\boldsymbol{q}_m$ need to be computed to obtain the degree $N$ polynomial surrogate expansion
\begin{equation}
\boldsymbol{x}(t,\boldsymbol{\theta}) \approx \boldsymbol{x}_N(t,\boldsymbol{\theta}) = \sum_{m=0}^{N} \boldsymbol{q}_m(t) \Phi_m(\boldsymbol{\theta}),
\end{equation}
which converges, for each vector entry, in $L^2_{\pi_\theta}(\mathcal{P})$ under mild assumptions \cite{ernst2012convergence}. 

In order to determine the coefficients $\boldsymbol{q}_m$, it is possible to employ an intrusive or a non-intrusive approach. We will focus on the intrusive approach, because of its superior accuracy and the strict orthogonal projection to the solution space. Again, we can insert the gPC expansion into the dynamical system to form the residual 
\begin{equation}
    \boldsymbol{r}_{\text{gPC}}(\dot{\boldsymbol{x}}_N,\boldsymbol{x}_N,t,\boldsymbol{\theta}) = \dot{\boldsymbol{x}}_N - \boldsymbol{f}(\boldsymbol{x}_N,t,\boldsymbol{\theta}).
    \label{eqn:residualgPC}
\end{equation}
Introducing the $L^2_{\pi_{\boldsymbol{\theta}}}(\mathcal{P})$ inner product as
\begin{equation}
    \left\langle \Phi_m , \Phi_n \right\rangle_{L^2_{\pi_{\boldsymbol{\theta}}}(\mathcal{P})} = \int_{\mathcal{P}} \Phi_m(\boldsymbol{\theta}) \Phi_n(\boldsymbol{\theta}) \pi_{\boldsymbol{\theta}}(\boldsymbol{\theta}) d\boldsymbol{\theta},
\end{equation}
we formulate the intrusive Galerkin approach to determine the gPC coefficients as
\begin{equation}
     \left\langle r_{\text{gPC},c}(\dot{\boldsymbol{x}}_N,\boldsymbol{x}_N,t,\cdot), \Phi_n \right\rangle_{L^2_{\pi_{\boldsymbol{\theta}}}(\mathcal{P})} = \int_{\mathcal{P}} r_{\text{gPC},c}(\dot{\boldsymbol{x}}_N,\boldsymbol{x}_N,t,\boldsymbol{\theta}) \Phi_n(\boldsymbol{\theta}) \pi_{\boldsymbol{\theta}}(\boldsymbol{\theta}) d \boldsymbol{\theta} = 0,
    \label{eqn:innerProductgPC}
\end{equation}
for $n=0,\ldots,N$.

\subsection{Combined Fourier-gPC analysis}
\label{sec:explainedFgPC}

In order to analyze uncertainties in periodic solutions of system \eqref{eqn:systemEquation}, we can apply the gPC approximation directly to the Fourier coefficients of the HB method. In particular, we can write 
\begin{equation}
    \boldsymbol{x}_{HN}(t,\boldsymbol{\theta}) = \sum_{k =-H}^{H} \sum_{m=0}^{N} \boldsymbol{q}_{km} e^{i k \omega t} \Phi_m(\boldsymbol{\theta}),
    \label{eqn:solutionFgPC}
\end{equation}
which is a combined Fourier-gPC surrogate. Inserting \eqref{eqn:solutionFgPC} into \eqref{eqn:systemEquation} results in the residual 
\begin{equation}
    \boldsymbol{r}_{\text{HN}}\left(\dot{\boldsymbol{x}}_{HN}, \boldsymbol{x}_{HN}, t, \boldsymbol{\theta} \right) = \dot{\boldsymbol{x}}_{HN} - \boldsymbol{f}(\boldsymbol{x}_{HN},t,\boldsymbol{\theta}).
    \label{eqn:residualFgPC}
\end{equation}
Since $x_{HN,i}$, $i=1,\ldots,n_{\text{d}}$, now lives in the product space $L^2([0,T]) \times L^2_{\pi_{\boldsymbol{\theta}}}(\mathcal{P})$, we introduce a new inner product as 
\begin{equation}
    \langle \langle u,v \rangle \rangle = \frac{1}{T}  \int_{\mathcal{P}} \int_0^T u(t,\boldsymbol{\theta}) \overline{v}(t,\boldsymbol{\theta}) \ d t \ \pi_{\boldsymbol{\theta}}(\boldsymbol{\theta}) d \boldsymbol{\theta} .
    \label{eqn:DefinnerProductFgPC}
\end{equation}
Again, a Galerkin projection leads to 
\begin{equation}
    \langle \langle r_{\text{HB},c}\left(\dot{\boldsymbol{x}}_{HN}, \boldsymbol{x}_{HN}, t, \boldsymbol{\theta} \right), \Psi_l(t) \Phi_n(\boldsymbol{\theta}) \rangle \rangle = 0, 
    \label{eqn:innerProductFgPC}
\end{equation}
for $l=-H,\ldots,H$ and $n=0,\ldots,N$.

For the numerical calculation, we derive a matrix vector representation of the inner product \eqref{eqn:innerProductFgPC}. We use the right-hand side of \eqref{eqn:HBdetEq} of the AFTHB method for the inner product of the Fourier analysis as explained in Section \ref{sec:harmonicbalance}. In the FgPC setting, this is rewritten as 
\begin{equation}
    \Tilde{\boldsymbol{r}}_{\text{F}} \left(\boldsymbol{q},\boldsymbol{\theta} \right) = \boldsymbol{E}_{HN_{t}}^{*} \Tilde{\boldsymbol{r}}_{\text{HB}}\left(\left(\boldsymbol{E}_{N_{t}H} \omega \boldsymbol{H} \otimes \boldsymbol{I}_N \right) \boldsymbol{q}, \left(\boldsymbol{E}_{N_{t}H} \otimes \boldsymbol{I}_N \right) \boldsymbol{q}, \tilde{\boldsymbol{t}}, \boldsymbol{\theta} \right).
\end{equation}
Here, $\boldsymbol{q}$ contains the FgPC coefficients in the order given by \eqref{eqn:solutionFgPC}.
To approximate the inner product over the stochastic domain, we employ a weighted Gauss quadrature rule 
\begin{equation}
     \int_{\mathcal{P}} \Tilde{r}_{\text{F},c} \left(\boldsymbol{q},\boldsymbol{\theta} \right) \Phi_n(\boldsymbol{\theta}) \ \pi_{\boldsymbol{\theta}} d \boldsymbol{\theta} = \sum_{z=1}^{N_G} \Tilde{r}_{\text{F},c} \left(\boldsymbol{q},\boldsymbol{\theta}^{(z)} \right) \Phi_n(\boldsymbol{\theta}^{(z)}) w^{(z)},
\end{equation}
for $n=0,\ldots,N$.
Here, the residual and the polynomial are evaluated at the $N_G$ Gaussian quadrature points $\theta^{(z)}$, which we multiply with the quadrature weights $w^{(z)}$. Both the quadrature points and weights are chosen corresponding to the probability density function of the random variable \cite{xiu2010numerical}. To account for all $\Phi_n$ of the inner product, we use the Kronecker product and receive the final residual $\boldsymbol{r}_{\text{FgPC}}$, which results in the determining equations for all $\boldsymbol{q}_{km}$ if we set it equal to zero
\begin{equation}
     \boldsymbol{r}_{\text{FgPC}}= \sum_{z=1}^{N_G} \left( \Tilde{\boldsymbol{r}}_{\text{F}} \left(\boldsymbol{q},\boldsymbol{\theta}^{(z)} \right) \otimes \boldsymbol{\Phi}(\boldsymbol{\theta}^{(z)}) \right) w^{(z)} = \boldsymbol{0}.
    \label{eqn:FgPCdetEq}
\end{equation}
Vector $\boldsymbol{\Phi}$ contains all polynomials $\Phi_n$. Equations \eqref{eqn:FgPCdetEq} are prone to become very large as $H$ and $N$ increase - making this method susceptible to the curse of dimensionality. 

The roots of \eqref{eqn:FgPCdetEq} yield the FgPC coefficients $\boldsymbol{q}_{km}$. Therefore, a root-finding algorithm needs to be applied in order to solve for the coefficients. In our experiments, we observed that a good initial guess is crucial. We achieved this by performing a time integration at the nominal parameter vector and applying a FFT to extract initial values for the first gPC coefficients of each harmonic. It must be mentioned, that we receive the cosine $\boldsymbol{a}_{k0}$ and sine $\boldsymbol{b}_{k0}$ amplitudes from the FFT, which are then transformed into the complex domain using Euler's formula. Details are given in \cite{de_jong_2024_13359870}.

\begin{figure}[htpb]
    \centering
    \includegraphics{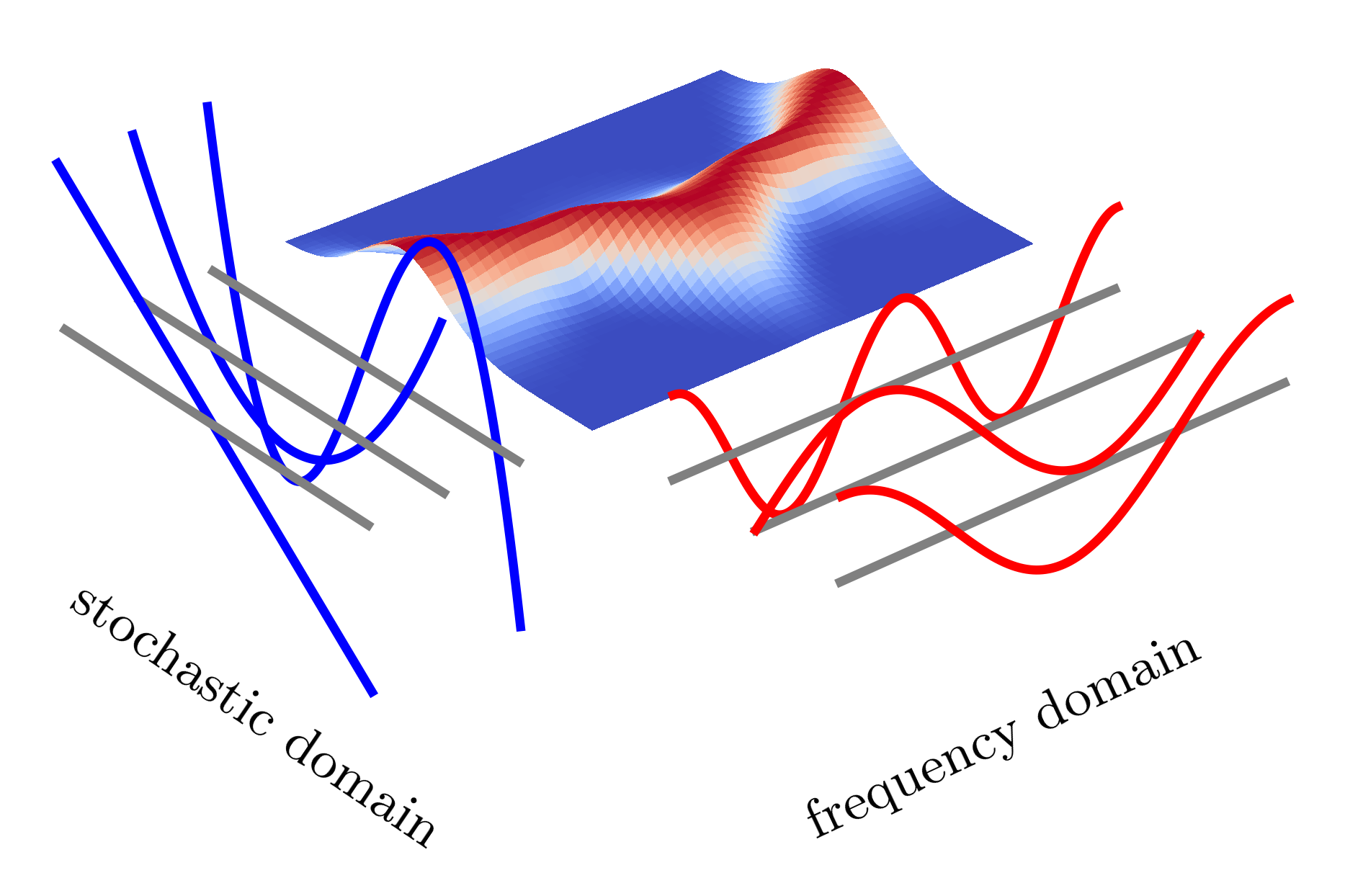}
    \caption{Visual representation of Fourier generalized Polynomial Chaos expansion}
    \label{fig:FgPCCoverpic}
\end{figure}

A visual representation of the aforementioned combination of the Fourier and polynomial series is presented in Figure \ref{fig:FgPCCoverpic}.

If the system is self-excited, the base frequency $\omega$ is an unknown and therefore approximated by a gPC surrogate model. Because the periodic solution is not associated with a fixed time, one phase value can be set arbitrarily and therefore reduces the number of unknowns \cite{trinhNonlinearDynamicBehaviour2020}. In our framework \cite{de_jong_2024_13359870}, we use the cosine and sine representation as input parameters, where $\boldsymbol{a}_{km}$ denotes the cosine and $\boldsymbol{b}_{km}$ the sine coefficient of the $k$-th amplitude and the $m$-th polynomial degree. The phase information is set with a fixed sine amplitude. We set the value of the sine amplitude of the first harmonic obtained by the FFT  to $\boldsymbol{b}_{10}$. The remaining gPC coefficients of the phase information $\boldsymbol{b}_{11}$ - $\boldsymbol{b}_{1N}$ are set to zero. With this, \eqref{eqn:solutionFgPC} changes to
\begin{equation}
\begin{aligned}
    \boldsymbol{x}_{HN}(t,\boldsymbol{\theta}) = &\left(\boldsymbol{q}_{-1,0} + \boldsymbol{q}_{1,0} \right) e^{i \sum_{m=0}^{N} q_{\omega m} \Phi_m(\boldsymbol{\theta}) t} \\
    & + \sum_{k =-H,k\neq -1,1}^{H} \sum_{m=0}^{N} \boldsymbol{q}_{km} e^{i k \sum_{m=0}^{N} q_{\omega m} \Phi_m(\boldsymbol{\theta}) t}\Phi_m(\boldsymbol{\theta}),
    \label{eqn:solutionFgPC_selfexcited}
\end{aligned}
\end{equation}
where $\boldsymbol{q}_{km}=\overline{\boldsymbol{q}}_{-k,m}$ are determined using Euler's formula $\boldsymbol{q}_{-k,m} = 0.5 ( \boldsymbol{a}_{km} \newline + i \boldsymbol{b}_{km})$.

\subsection{Deflation technique}
\label{sec:deflation}

As mentioned above, nonlinear systems can have multiple solutions for one parameter set. For the benchmark problem, the Duffing oscillator, this also holds true for the selected frequency parameter. Here, two stable and one unstable solutions are present, which we would like to evaluate for stochastic predictions. 
When applying the HB or FgPC method, each solution has a different basin of initial conditions for the root-finding algorithm. Therefore, it highly depends on the initial guess which solution is found. In order to find as many solutions as possible, the algorithm needs to be rerun with appropriate initial guesses. To make the algorithm more robust and capable of finding multiple solutions, it can be equipped with the deflation technique, which works well with the Newton-Raphson method. A good introduction is presented in \citet{farrellDeflationTechniquesFinding2015}. The fundamental premise is that a singularity is incorporated into the residual for each solution, thereby displacing the Newton-Raphson method from the identified solution. In some cases, an alternative initial estimate may be necessary to facilitate convergence towards an alternative solution. 

For our purpose, we use the so-called shifted deflation technique. For a solution $\boldsymbol{s}$, the deflation operator looks like
\begin{equation}
	\mathcal{D}_{\boldsymbol{s}}(\boldsymbol{q}) = \frac{\boldsymbol{I}_{\boldsymbol{q}}}{\|\boldsymbol{s}-\boldsymbol{q}\|^{p_{\text{D}}}} + \alpha_{\text{D}} \boldsymbol{I}.
	\label{eqn:deflationOperator}
\end{equation}
This operator combines all positive characteristics of the deflation technique presented in \cite{farrellDeflationTechniquesFinding2015}. 
$\boldsymbol{q}$ is the current point used by the Newton-Raphson method, $\boldsymbol{I}_{\boldsymbol{q}}$ the identity matrix of the same size as $\boldsymbol{q}$, and $p_{\text{D}}$ and $\alpha_{\text{D}}$ are deflation parameters. The deflation parameters need to be tuned according to the underlying problem. 

The first term of \eqref{eqn:deflationOperator} creates the singularity if the Newton-Raphson method chooses a point close to a found solution. The second one ensures that the modified function does not exhibit asymptotic behavior where the root algorithm might look for a solution.

In order to add the operator to the FgPC method, it is multiplied to the left-hand side of the vector constructed of all Galerkin projections of \eqref{eqn:innerProductFgPC}: $\mathcal{D}_{\boldsymbol{s}}(\boldsymbol{q}) \boldsymbol{r}_{\text{FgPC}}(\boldsymbol{q})$.
Each solution has its own operator - and they are multiplied with each other, e.g. for two solutions we obtain $\mathcal{D}_{\boldsymbol{s}_2}(\boldsymbol{q}) \mathcal{D}_{\boldsymbol{s}_1}(\boldsymbol{q}) \boldsymbol{r}_{\text{FgPC}}(\boldsymbol{q})$.

\section{Numerical Setup and Duffing Oscillator Benchmark}
\label{sec:duffing}

A code framework implementing the FgPC method described in this paper is available at Zenodo, see \citet{de_jong_2024_13359870}. We utilized the packages NumPy \cite{harris2020array}, SciPy \cite{2020SciPy-NMeth}, and Chaospy \cite{FEINBERG201546} for our computations. The computations were carried out on a workstation equipped with 24 cores and 126 GB of RAM.
This sets the stage for the numerical calculations conducted in this paper, which will be discussed next in the context of a simple, academic benchmark problem: the Duffing oscillator
\begin{equation}
    \ddot{x} + \delta \dot{x} + \alpha x + \beta x^3 = \gamma \cos{\Omega t}, x(0) = x_0.
    \label{eqn:duffing}
\end{equation}

All calculations are based on the settings in Table \ref{tab:DuffingParameters}, which are mass normalized. 
\begin{table}[htbp]
    \centering
    \caption{Duffing oscillator parameter values}
    \label{tab:DuffingParameters}
    \begin{tabular}{cccc}
        \hline
        $\delta = \qty{0.08}{\per\second} $ \rule{0pt}{3ex} & $ \beta = \qty{1}{\per\square\second\per\square\meter}$ & $\gamma = \qty{0.2}{\meter\per\square\second}$ & $\Omega = \qty{1.4}{\hertz}$ \\
        \hline
    \end{tabular}
\end{table}
The linear stiffness is considered uncertain with $\alpha \sim \text{Beta}(5,5,0.8, \newline 1.2)$. We chose the Beta distribution since it features a bounded support - here with a value in the range of $[0.8, 1.2]$.

\begin{figure}[htbp]
    \centering
    \includegraphics{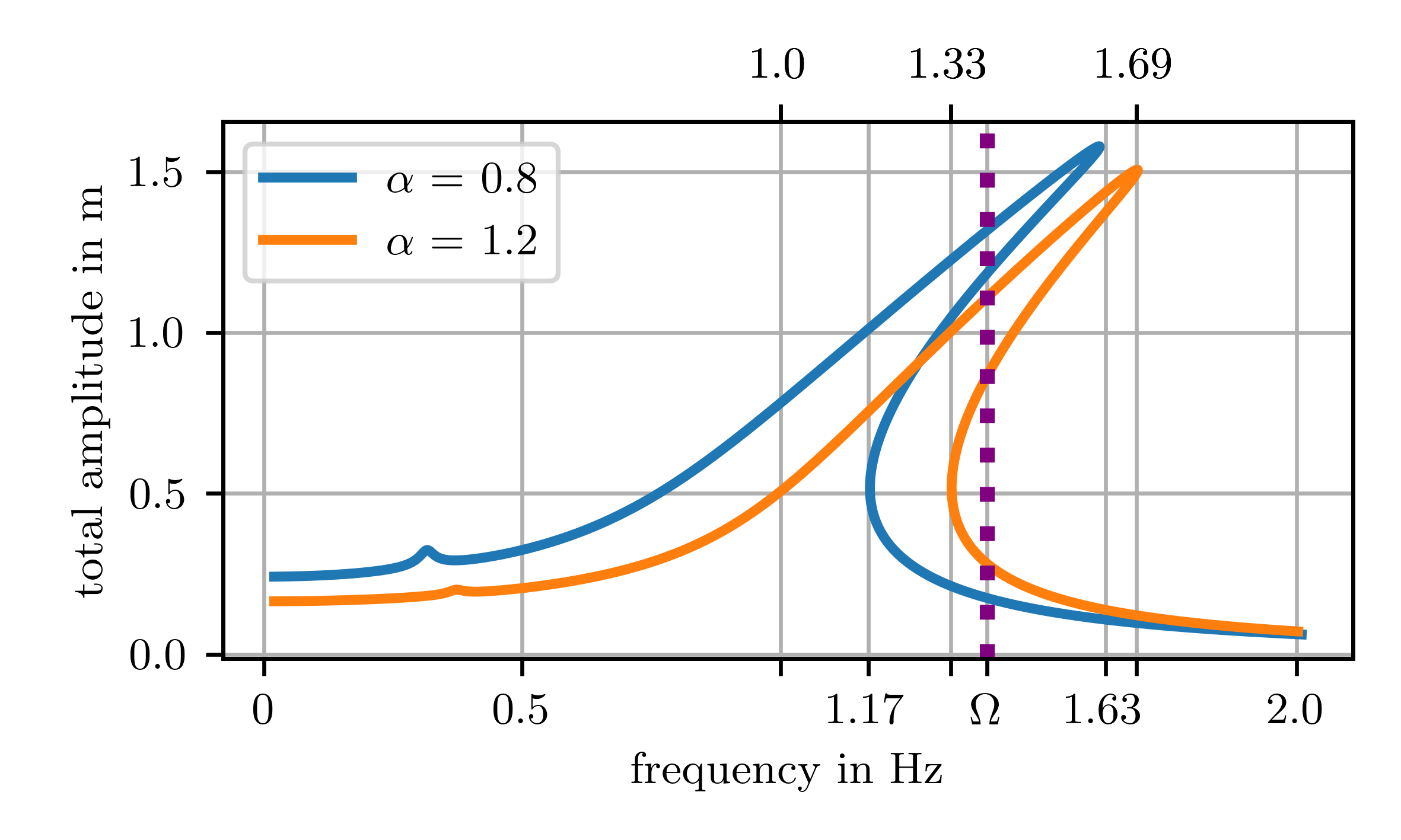}
    \caption{Backbone curve of the Duffing oscillator for different values of $\alpha$}
    \label{fig:backboneDuffing}
\end{figure}

For both boundary points, the frequency response functions are depicted in Figure \ref{fig:backboneDuffing}. They represent the typical backbone curve of the Duffing oscillator. 
The curves are determined through the HB in conjunction with a path continuation via a predictor-corrector step, as outlined in \cite{krack2019}. Due to the range in the linear stiffness, a horizontal shift in the backbone curve occurs. In particular, this results in the Duffing oscillator having either one or three solutions for the frequency ranges $\Omega \approx [1.17, 1.33]$ and $[1.63, 1.69]$. There are three solutions in the range $\Omega \approx [1.33, 1.62]$, and only one in the remaining range of $\Omega$. With the indicated parameters, we find ourselves in a setting with three solutions. Here, we are able to analyze the stochastic properties of each solution individually with the FgPC method. In future work, we will address a stochastic description of all solutions at the same time.

In order to apply the FgPC method to the Duffing oscillator, \eqref{eqn:solutionFgPC} is differentiated twice in time and incorporated into the residual form of the Duffing oscillator
\begin{equation}
	r_{\text{Duf}} = \ddot{x}_{HN} + \delta \dot{x}_{HN} + \theta x + \beta x_{HN}^3 - \gamma \cos{\Omega t} \text{.}
	\label{eqn:residuumDuffing}
\end{equation}
This residual is used in \eqref{eqn:innerProductgPC} to derive the determining equations. An initial guess is drawn as stated in Section \ref{sec:FgPC}. This initial guess is then used to find the root with the modified Powell's method \textit{hybr} from the Scipy optimization's \textit{root} function. Using the deflation technique, it is possible to find all three solutions. Only one initial guess is needed when $N = 0$. Otherwise, a new initial guess for each solution is required.

\begin{figure}[htb]
    \centering
    \includegraphics{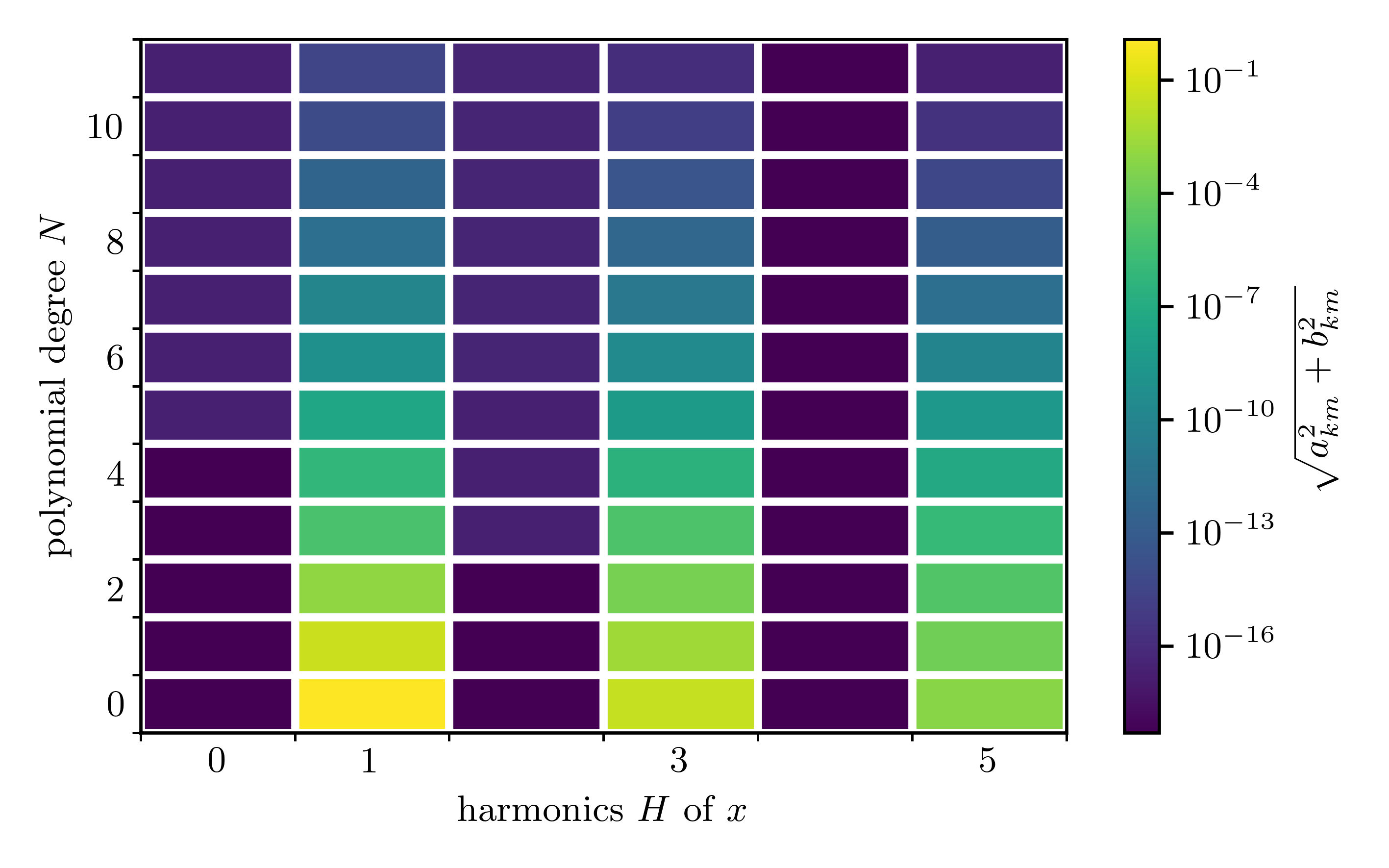}
    \caption{Magnitude of each coefficient for every combination of the harmonic $H$ and polynomial degree $N$ for poition $x$ of the Duffing oscillator}
    \label{fig:coeffGridDuffing}
\end{figure}

In Figure \ref{fig:coeffGridDuffing}, the coefficients are presented in the following manner: The determined coefficients, denoted as $q_{km}$, are received in cosine $a_{km}$ and sine $b_{km}$ representation from the algorithmic process. These representations serve to ascertain the norm of each harmonic and polynomial degree combination, which is expressed as the square root of the sum of the squares of the coefficients, represented as $\sqrt{a_{km}^2+b_{km}^2}$. 

As anticipated, the magnitude declines with the increase in harmonics and polynomial degree, pointing out the potential for an adaptive sparse approach to FgPC. In the context of the Duffing oscillator, this approach could significantly reduce the number of coefficients, as the constant component of the Fourier series and even harmonics are absent.

To further validate the truncation scheme, we conducted a convergence study, employing the following root mean square error as a metric: 
\begin{equation}
     \varepsilon_{HN} = \sqrt{\frac{1}{N_s} \sum_{j=0}^{N_s}   \left(\frac{1}{T} \int_0^T | x_{HN}(\theta_j,t) | dt- \frac{1}{T} \int_0^T |x_{\text{ref}}(\theta_j,t) | dt \right)^2 },
    \label{eqn:relError}
\end{equation}
where $x_\text{ref}$ is the solution with the highest harmonic $H$ and polynomial degree $N$, and $N_s$ indicates the number of samples used. Each solution is integrated over one period so as to avoid phase shifts between the solutions, which are especially likely to occur when the system is self-excited. 

\begin{figure}[htb]
    \centering
\includegraphics{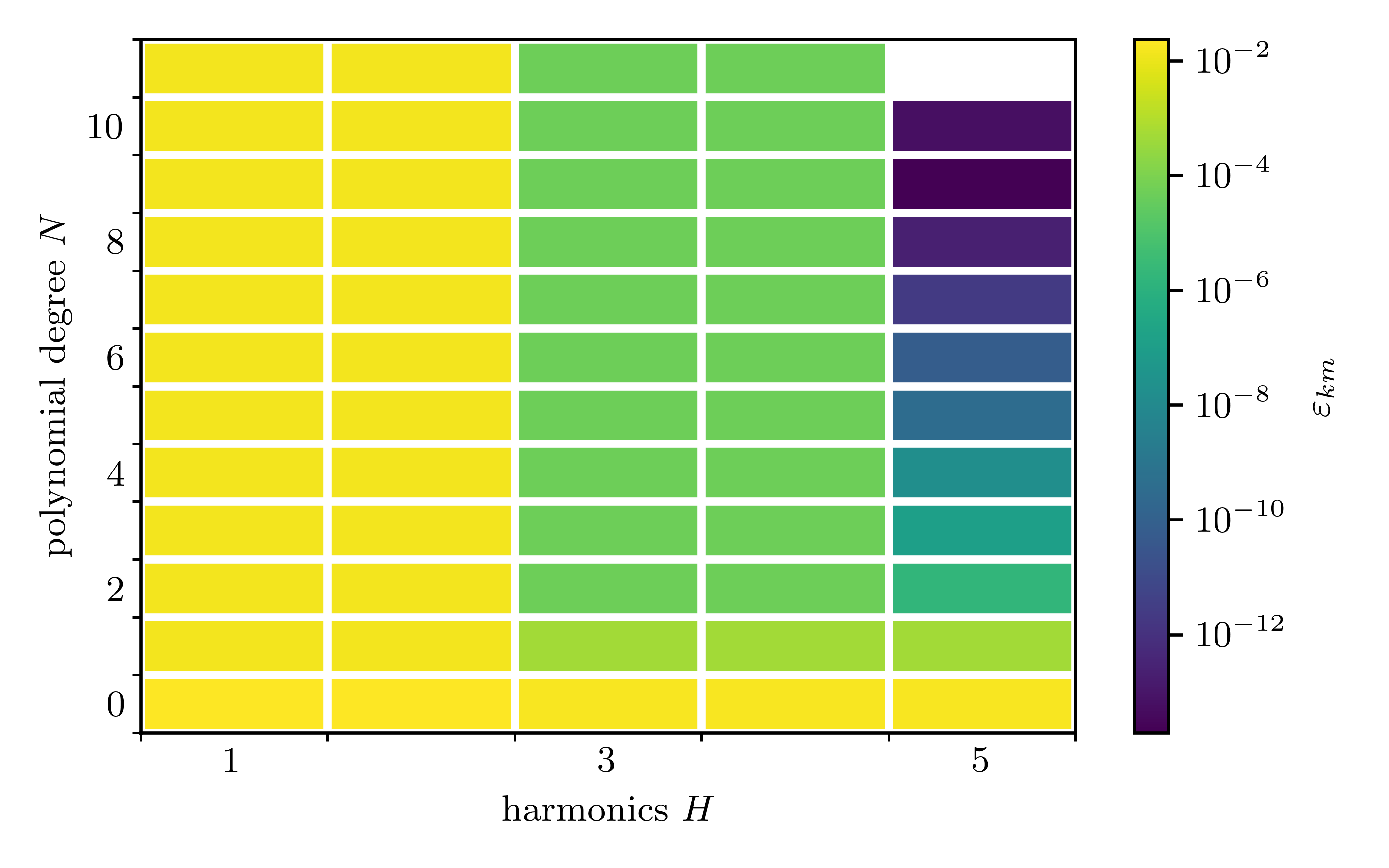}
    \caption{Relative error of the FgPC method for every combination between the harmonic $H$ and polynomial degree $N$ for the Duffing oscillator}
    \label{fig:convergenceDuffing}
\end{figure}

As illustrated in the coefficient grid of Figure \ref{fig:coeffGridDuffing}, it is sufficient to consider only odd harmonics. This behavior is also evident in the convergence study presented in Figure \ref{fig:convergenceDuffing}. In this instance, the reference solution is designated as $x_{5,11}$. As even harmonics do not contribute to further accuracy, the error remains the same with the previous odd harmonic. This observation allows us to conclude that, for lower harmonics, the error does not decrease further as the polynomial degree increases above two. However, since we are measuring a relative error, it contributes less than solutions with higher harmonics. 

The coefficient grid and the convergence study exhibit a high degree of alignment, whereby an increase in both harmonics and polynomial degree is accompanied by a reduction in both the relative error and coefficient magnitude. Accordingly, the FgPC methodology can be employed to ensure a well-suited FgPC surrogate model.

\begin{figure}[h]
	\centering
	\begin{subfigure}{0.48\textwidth}
		\centering
		\includegraphics{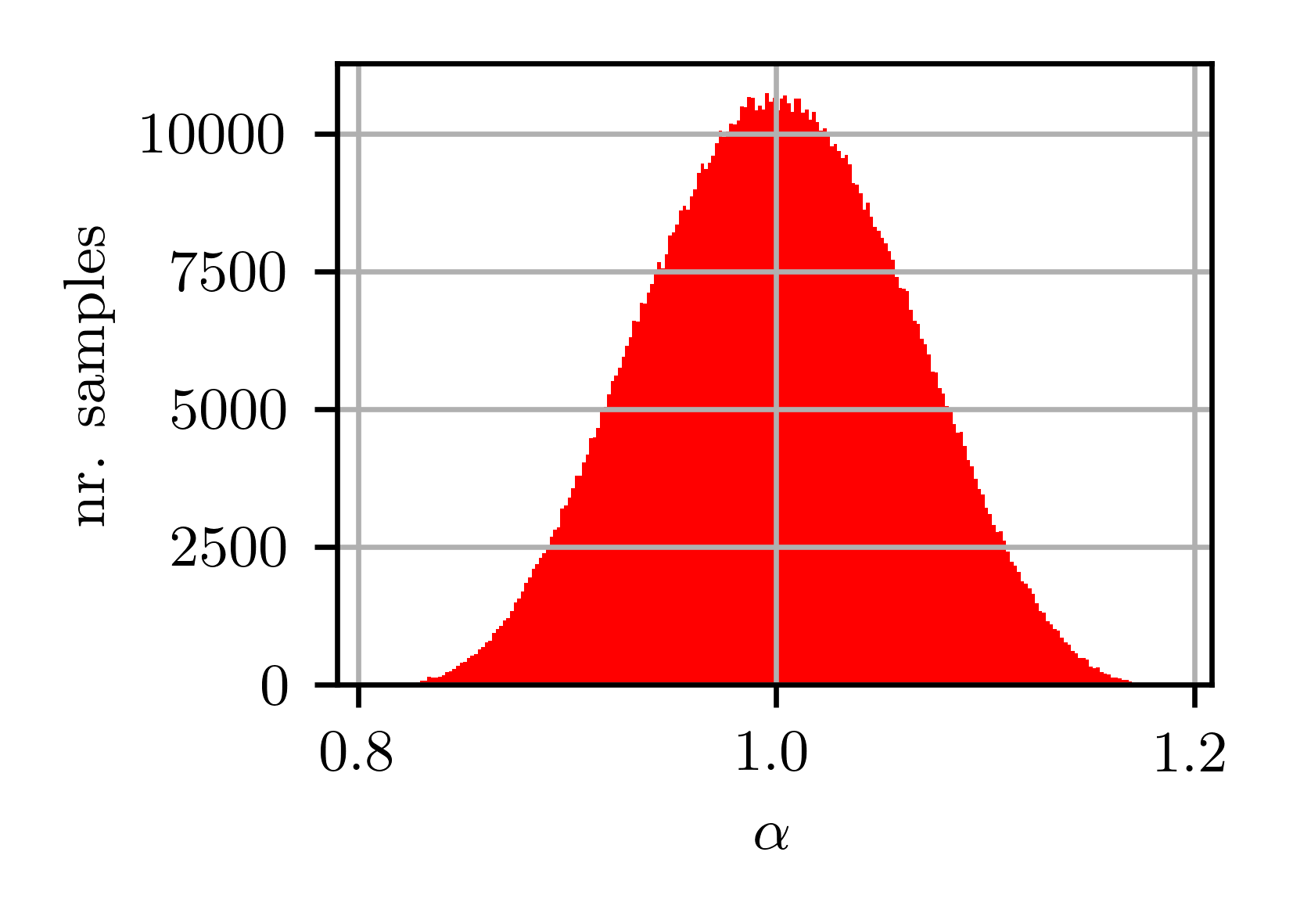}
		\caption{Beta distribution of linear stiffness}
		\label{fig:inputDist}
	\end{subfigure}
	\hfill
	\begin{subfigure}{0.48\textwidth}
		\centering
		\includegraphics{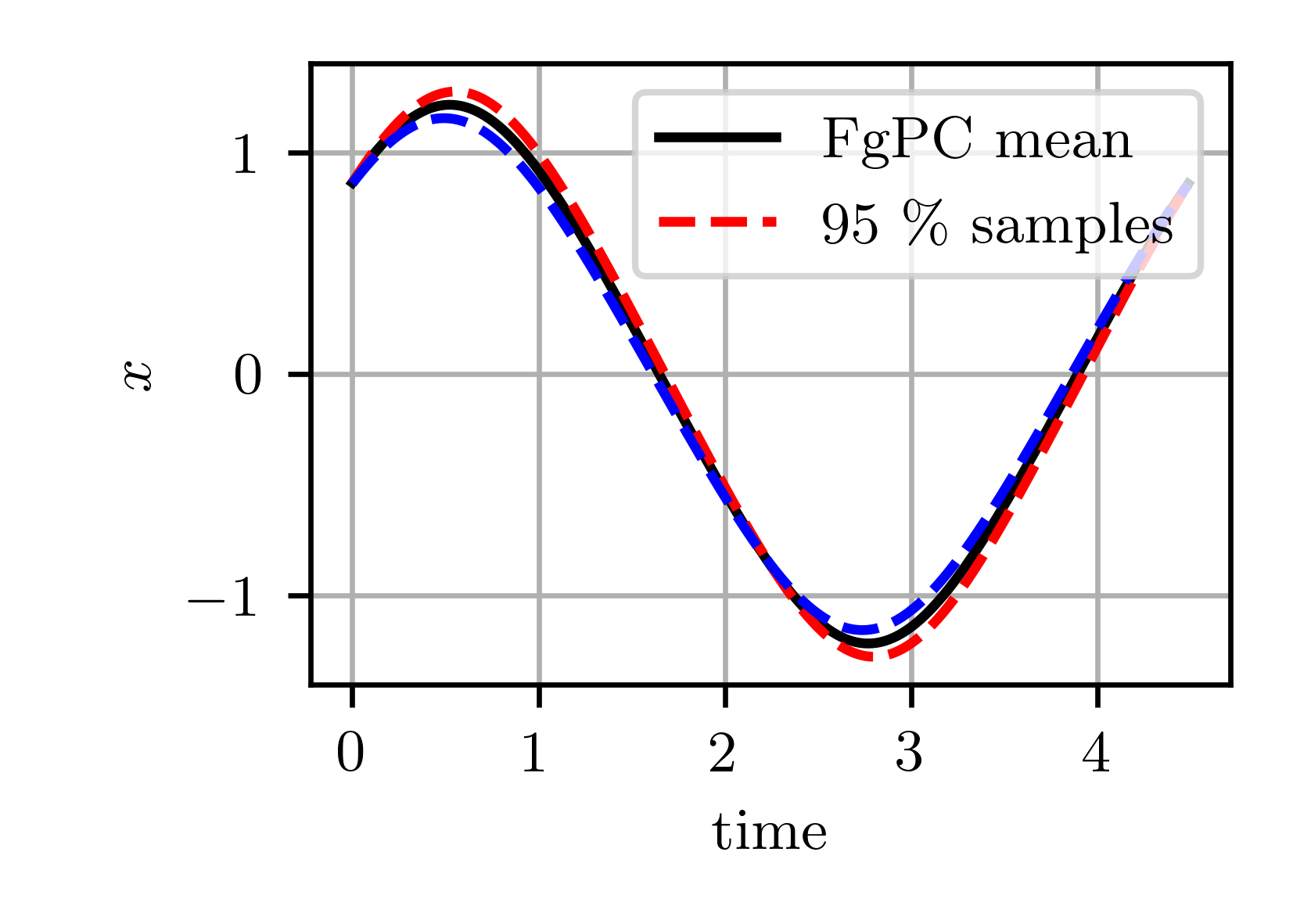}
		\caption{Mean value and interval containing 95 \% of all samples of the position over one period }
		\label{fig:duffingPosition}
	\end{subfigure}
	\par \medskip
	\centering
	\begin{subfigure}{0.48\textwidth}
		\centering
		\includegraphics{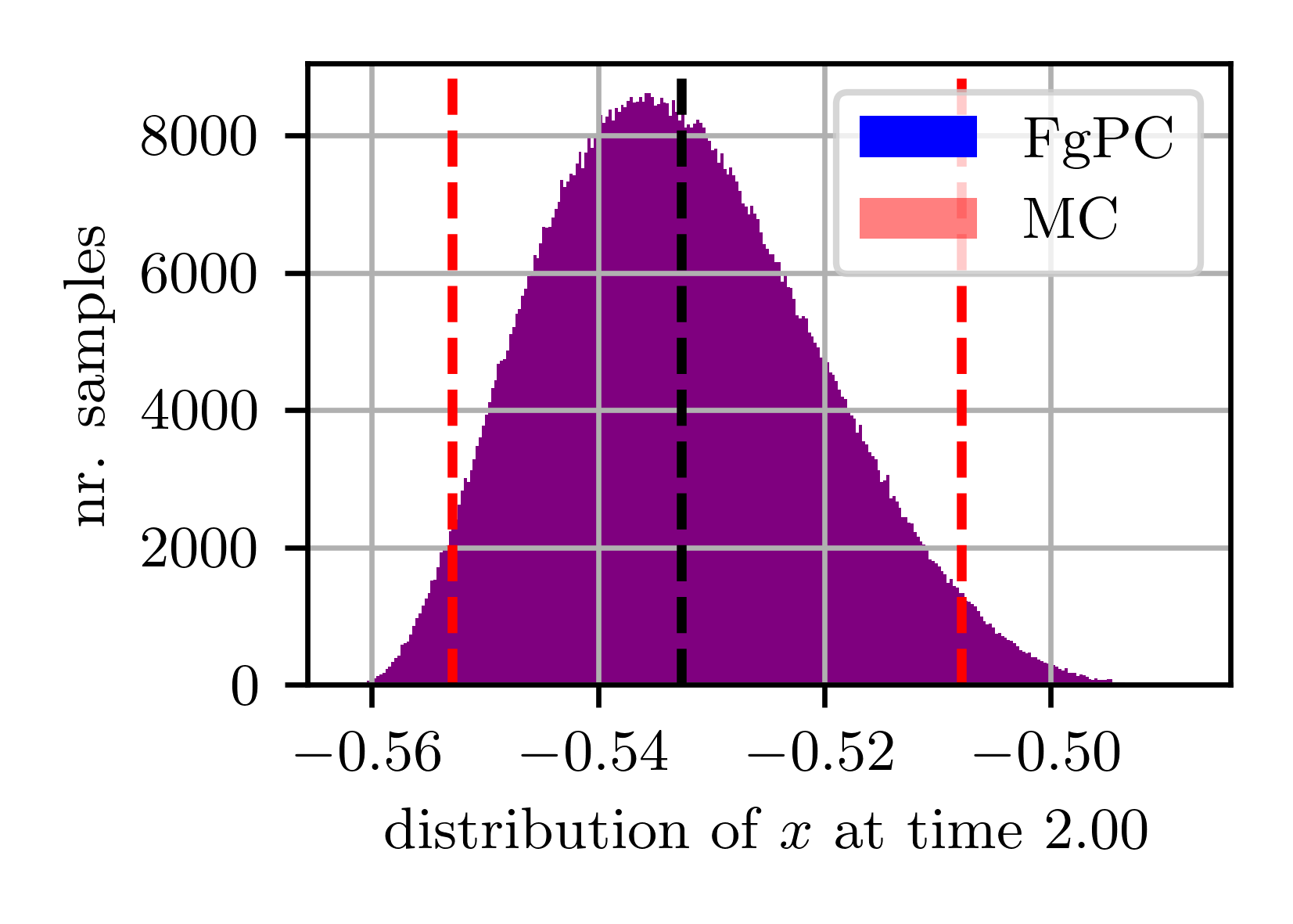}
		\caption{Distribution at time point \qty{2}{\second} of the FgPC and MC method, with the mean value (black dashed line) and the interval containing 95 \% of all samples (red dashed line)}
		\label{fig:duffingDist_0}
	\end{subfigure}
	\hfill
	\begin{subfigure}{0.48\textwidth}
		\centering
		\includegraphics{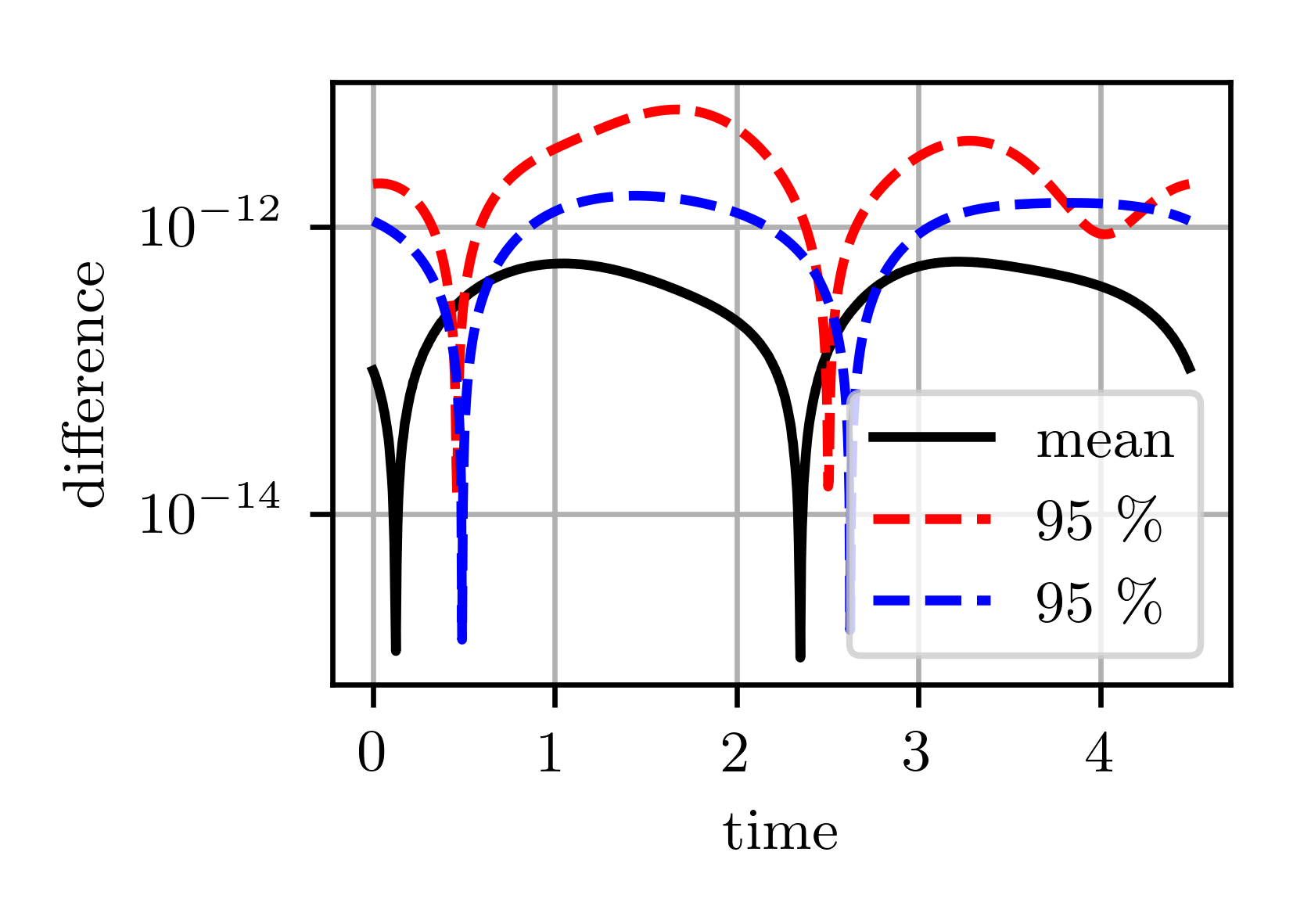}
		\caption{Difference between the FgPC and MC method for the mean value and the interval containing 95 \% of all samples of the position over one period}
		\label{fig:duffingDifference}
	\end{subfigure}
	\caption{Large stable FgPC solution of the Duffing oscillator with $H=5$ and $N=12$}
	\label{fig:duffingOscillator}
\end{figure}

Once the FgPC coefficients have been determined, it is possible to proceed with the calculation of the stochastic response of the Duffing oscillator. To this end, we draw $1e6$ samples from the input Beta distribution for the linear stiffness. The sample distribution is illustrated in Figure \ref{fig:inputDist}. 
The aforementioned distribution is then subjected to the FgPC method, resulting in a random process of the system variable over the course of one period, which is approximately $T = \frac{2\pi}{\Omega} \approx \qty{4.49}{\second}$. This allows us to capture the marginal distributions for given time points within the period. Figure \ref{fig:duffingPosition} depicts the time process for the large solution of the Duffing oscillator, including the mean value and a 95 \% sample coverage interval. This interval encompasses the two samples that are situated at the periphery of the centered 95 \% interval of the samples at the specified time point. As the input distribution is ordered from the smallest to the greatest value and passed through the FgPC in this sequence, the red dashed line represents the sample closest to 2.5 \% of all samples, while the blue line depicts the sample closest to 97.5 \% of all samples. This is consistent with the backbone curve in Figure \ref{fig:backboneDuffing}, which illustrates that the magnitude of the total amplitude of an $\alpha$ close to $0.8$ is greater than that of an $\alpha$ close to $1.2$. Consequently, the red boundary has a greater magnitude, causing the blue and red boundaries to intersect twice within one period.

In order to facilitate a comparative analysis of the marginal distributions generated by the FgPC, a MC model was implemented as well. The MC model represents the HB method evaluated for each sample. Upon examination of the distribution at the time point $\qty{2}{s}$, we observe that the FgPC and MC method yielded identical distributions, as illustrated in Figure \ref{fig:duffingDist_0}. The distributions are asymmetrical, indicating the absence of a linear correlation between the input and output distributions.

Furthermore, a comparison can be made between the mean and the coverage interval boundaries of both methods over one period. The discrepancy between the two methods is illustrated in Figure \ref{fig:duffingDifference}. By employing the coefficients of a gPC expansion (see Section \ref{sec:gPC}), we can directly calculate stochastic moments, such as the mean value $\mathbb{E}[x_{HN}(t,\cdot)] = q_0(t)$ and the variance $\mathbb{V}[x_{HN}(t,\cdot)]=\sum_{m=1}^N q_m(t)^2$. Thus, the mean value is in excellent alignment even if a lower polynomial degree is chosen.

Even for this relatively simple example, the execution time for the FgPC method is considerably shorter than that of the MC method, at $\qty{32.74}{\minute}$ versus $\qty{552.93}{\minute}$, respectively. Therefore, for problems of increasing complexity, the advantages of the FgPC method are supposed to be even greater, as long as the dimension $p$ does not become too large. The aforementioned time periods take into account the calculation of the FgPC coefficients and  sequential sampling. The MC method requires the determination of the roots of the HB method. In this particular instance, sorting the sample is beneficial, as the solution for the roots of the previous sample can be employed as an initial guess for the next sample. 

\begin{figure}[htb]
    \centering
    \includegraphics{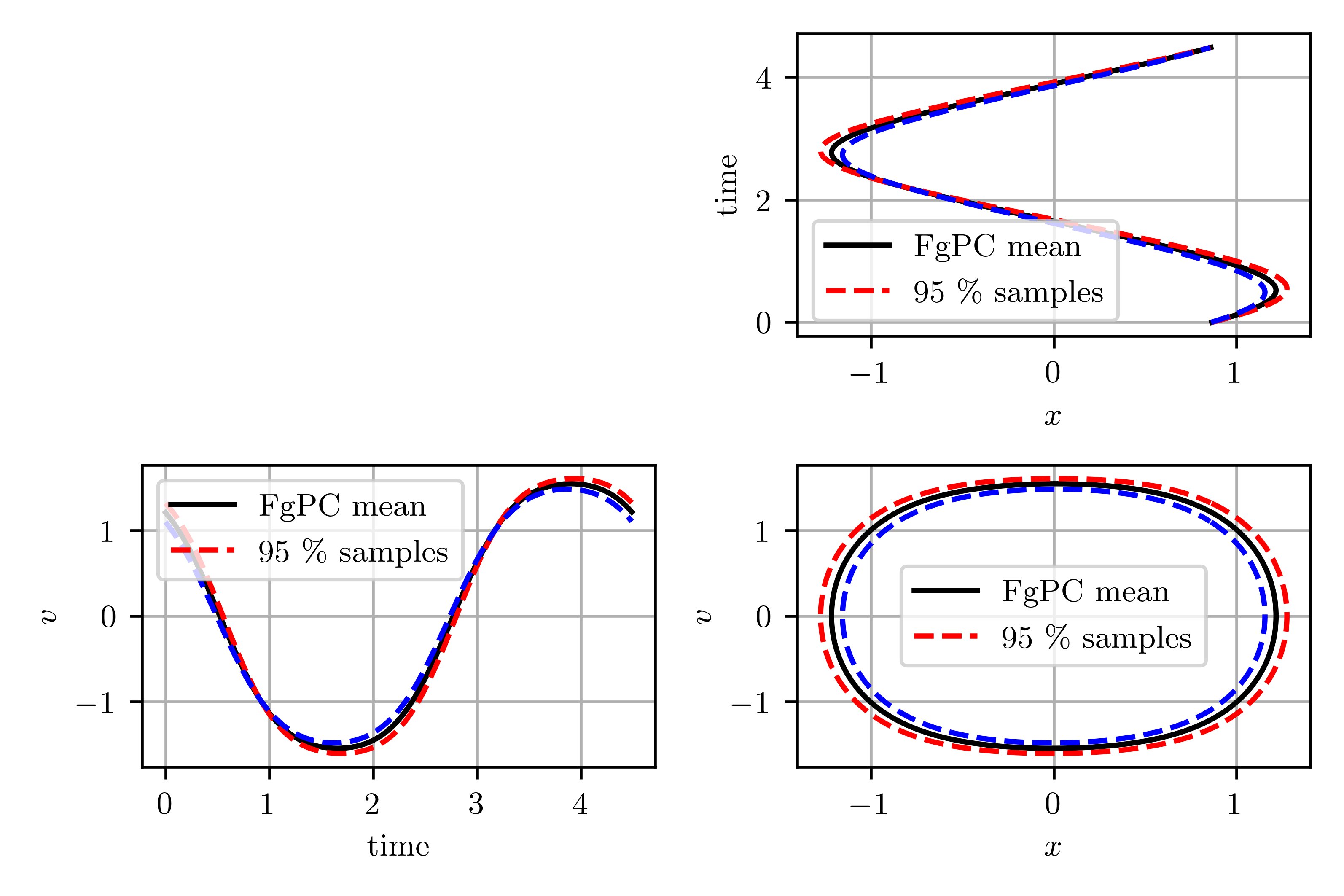}
    \caption{Phase portrait of the mean value and the interval containing 95 \% of all samples for the Duffing oscillator}
    \label{fig:duffingPhaseportrait}
\end{figure}

Given the uncertain time-dependent amplitude, we are able to construct the phase portrait of the Duffing oscillator, as illustrated in Figure \ref{fig:duffingPhaseportrait}. Of particular interest are the three circle-like closed curves, which are a consequence of the manner in which the coverage sample interval is determined. As previously described, the samples are ordered in a specific sequence, resulting in a reduction in the size of the phase portrait.

\section{Application: Electrophysiology of \texorpdfstring{$\beta$}{ß} Cells}
\label{sec:betaCell}

Having previously elucidated the FgPC method in the context of the benchmark problem, the Duffing oscillator, we can now apply it to a more complex model, describing the electrophysiology of $\beta$ cells. 
The $\beta$ cells are located in the pancreas and are responsible for the secretion of insulin, which is essential for regulating blood sugar levels in the human body. A dysfunction of the cells is often associated with the development of disease, as exemplified by type 2 diabetes. The prevalence of type 2 diabetes is increasing globally, and its treatment is of significant social and economic importance. However, it is inherent to biological systems that they are subject to a number of uncertainties, including those related to cell size and shape, distribution of organelles within the cell, age, and cell functionality \cite{gausChangingViewInsulin2022}. Accounting for these uncertainties and quantifying their impact on the behavior of the $\beta$ cells has the potential to inform more effective treatment strategies, particularly those tailored to the individual.  

\begin{figure}[htb]
	\centering
	\includegraphics[width = \textwidth]{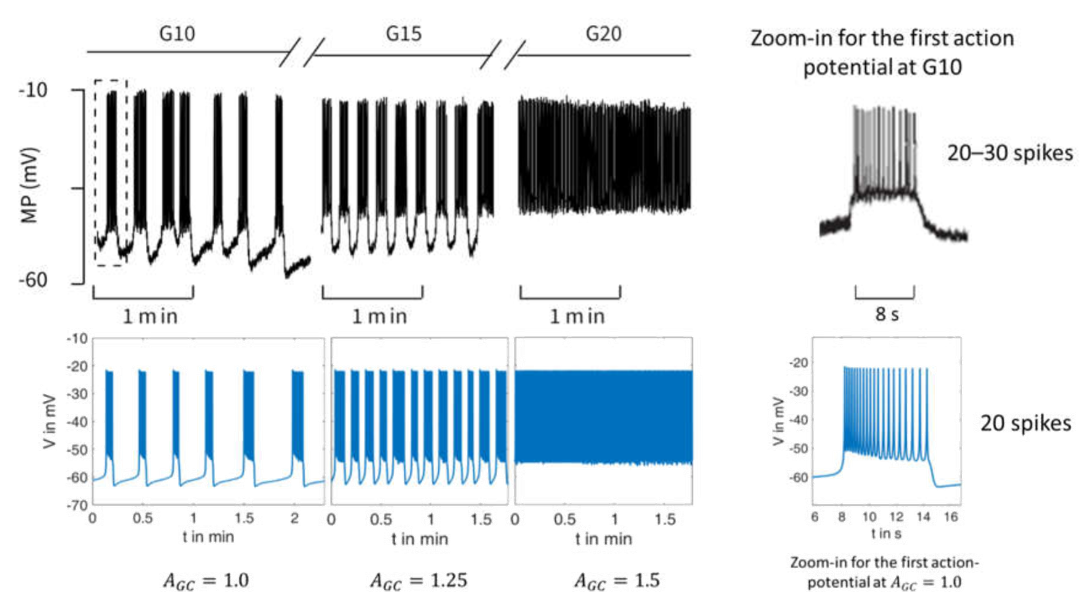}
	\caption{Experiment (upper half. Reprinted with permission from \cite{gilonCalciumSignalingPancreatic2014}, 2014, Elsevier) and simulation (lower half. For more information, please refer to \citet{mullerDynamicsCalciumSignaling2023}) of membrane potential for different extracellular glucose concentrations.}
	\label{fig:cellExpSim}
\end{figure}

From a biological standpoint, understanding the conditions under which periodic processes, particularly action potentials, can occur is of paramount importance. This is also the case with the $\beta$ cell. It is the oscillations of the electrical membrane potential that are of vital importance, as they control processes that are similar to the transmission of stimuli in nerve cells. These pulses regulate insulin secretion. In the absence of these processes, the cell is unable to function correctly. 
The aforementioned oscillations are illustrated in Figure \ref{fig:cellExpSim}. 
The figure presents a comparison of experimental data (upper half) with simulation data (lower half) of the membrane potential of the $\beta$ cell for different extracellular glucose concentrations. The simulation model demonstrates a high degree of alignment with the experimental data. As the glucose concentration increases, the alternation of bursts and rest phases transitions into a state where only bursts are observed. In the simulation model, the glucose concentration is regulated through the parameter $A_{\text{GC}}$. Upon closer examination of the bursts, a discernible change in the peaks and therefore in the period of the burst becomes evident. These bursts warrant particular attention. 

\begin{figure}[htb]
	\centering
	\includegraphics[width = 0.7 \textwidth]{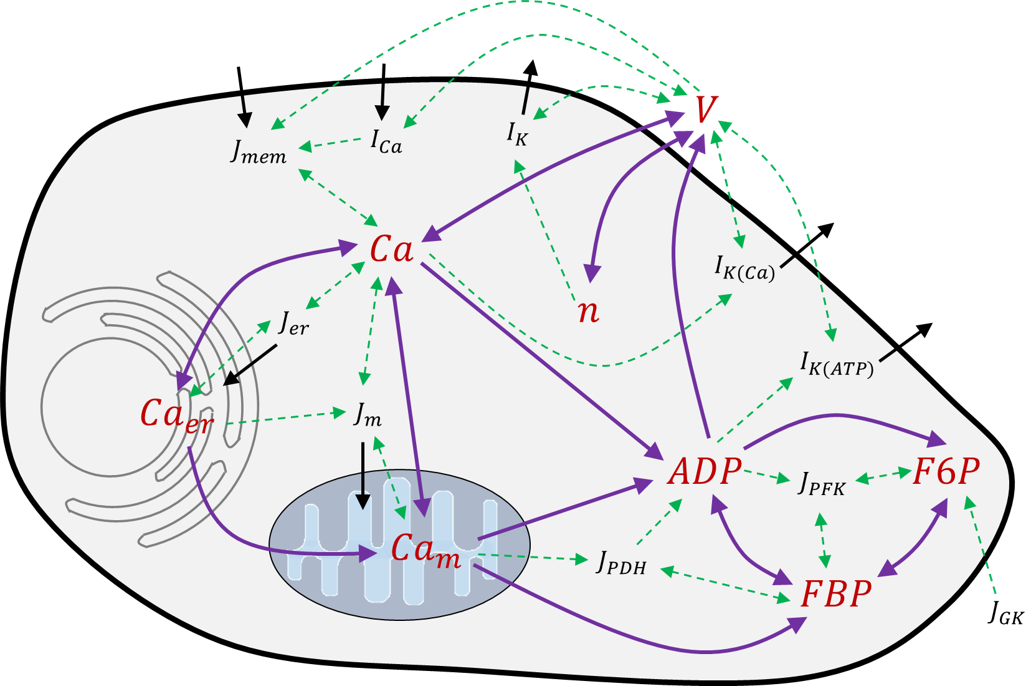}
	\caption{Schematic sketch of the IOM of a $\beta$ cell. The red letters are the governing eight state variables. The arrows describe interactions and dependencies. For more information, please refer to \citet{mullerDynamicsCalciumSignaling2023}}
	\label{fig:cellModel}
\end{figure}

The simulation model was developed over time based on empirical observations and established biological reactions. The model is called "integrated oscillator model" (IOM), and a schematic representation of the IOM of a $\beta$ cell is provided in Figure \ref{fig:cellModel} \cite{marinelliTransitionsBurstingModes2018}. 
The diagram illustrates the three primary phenomena occurring within the IOM. The initial phenomenon is the electrical potential of the cell membrane, denoted by $V$, which is subject to differing concentrations of ions inside and outside the cell. Secondly, there are diffusive currents (depicted by black arrows) which facilitate the equalization of ion concentration differences within and outside the cell. This is achieved through the controlled opening and closing of channels in the membrane, which are regulated by the concentration of specific ions, such as potassium or calcium (red letters). The green arrows indicate the explicit mathematical interactions of these channels with the state variables. Thirdly, violet arrows describe the coupling of the state variables, which refer to different processes including ion fluxes through the channels, diffusion processes, and chemical reactions.  

\begin{figure}[htb]
	\centering
	\includegraphics{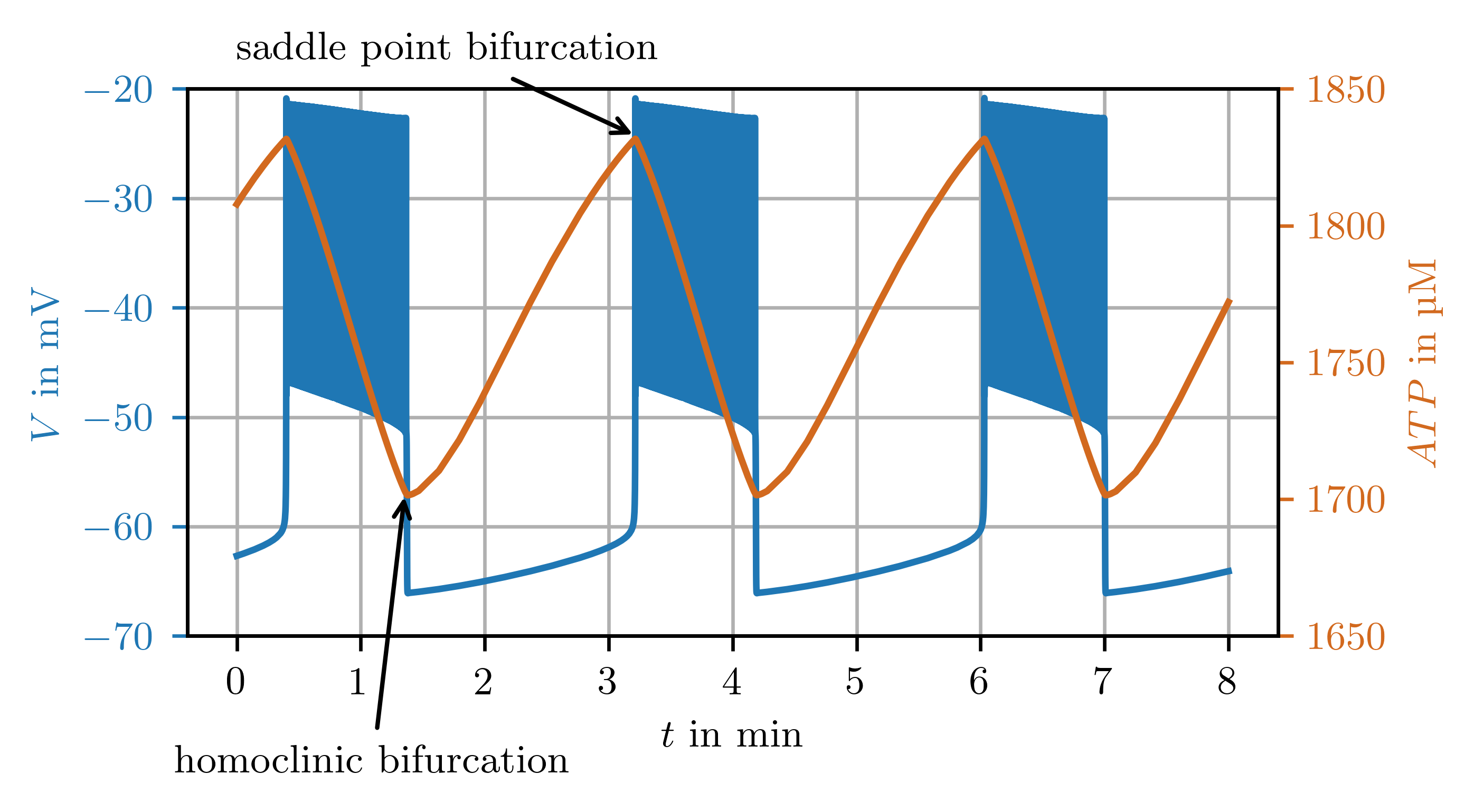}
	\caption{Resulting quasi-stationary state of the membrane potential $V$ and $ATP$ as determined by time integration of the IOM model}
	\label{fig:cellTimeint}
\end{figure}

It is important to note that the IOM is a multiscale model that exhibits oscillatory behavior. The process of glucose metabolism into $ATP$ (adenosine triphosphate) occurs on a relatively slow timescale, with periods of minutes. It is therefore classified as the metabolic system. In contrast, the membrane potential exhibits behavior on a fast scale, with periods of fractions of a second. It is consequently designated as the electrical system. The two systems are interconnected through $ATP$, which is in turn determined by $ADP$ (adenosine diphosphate). The aforementioned oscillations are illustrated in Figure \ref{fig:cellTimeint} as an exemplar of the membrane potential ($V$) and the $ATP$ concentration, which were calculated using a time integration method.
It is evident that $ATP$ plays a pivotal role in governing the primary characteristics of the electrical system, which we intend to examine in greater detail. This, along with the crucial function of the electrical oscillations within the cell, provides the rationale behind our initial focus on the electrical system. Because the $ATP$ value changes only marginally during one limit cycle of the electrical system, the isolated electrical system considered in the following studies assumes the value of $ATP$ as a fixed value. The governing equations of the electrical system are therefore of the form
\begin{align}
    \frac{dV}{dt} &= -\frac{1}{C} \left[ I_1(V) + I_2(V,n) + I_3(Ca,V) + I_4(V,ATP) \right] \label{eqn:electricV} \\
    \frac{dn}{dt} &= \frac{g(V)-n}{A} \label{eqn:electricn} \\
    \frac{dCa}{dt} &= B \left( J_1(V,Ca) - J_2(Ca)\right) \text{,}
    \label{eqn:electricc}
\end{align}
where the state variables are the membrane potential $V$, the activation function $n$ of the delayed rectifying $K^{+}$ current, and the intracellular $Ca^{2+}$ concentration $Ca$. The system functions $I_i$, $J_i$ and $g$ are also involved. $A$, $B$, and $C$ are system parameters. The system functions $I_i$ are interpreted as permeability functions depending on $V$, which change from zero to one within a sharply defined interval and have forms like $1/(1+exp((\nu- V)/s))$, where $\nu$ and $s$ are shape parameters. For a comprehensive and more detailed description of the model and its equations, please refer to \citet{marinelliTransitionsBurstingModes2018}.

\begin{figure}[htb]
	\centering
	\includegraphics{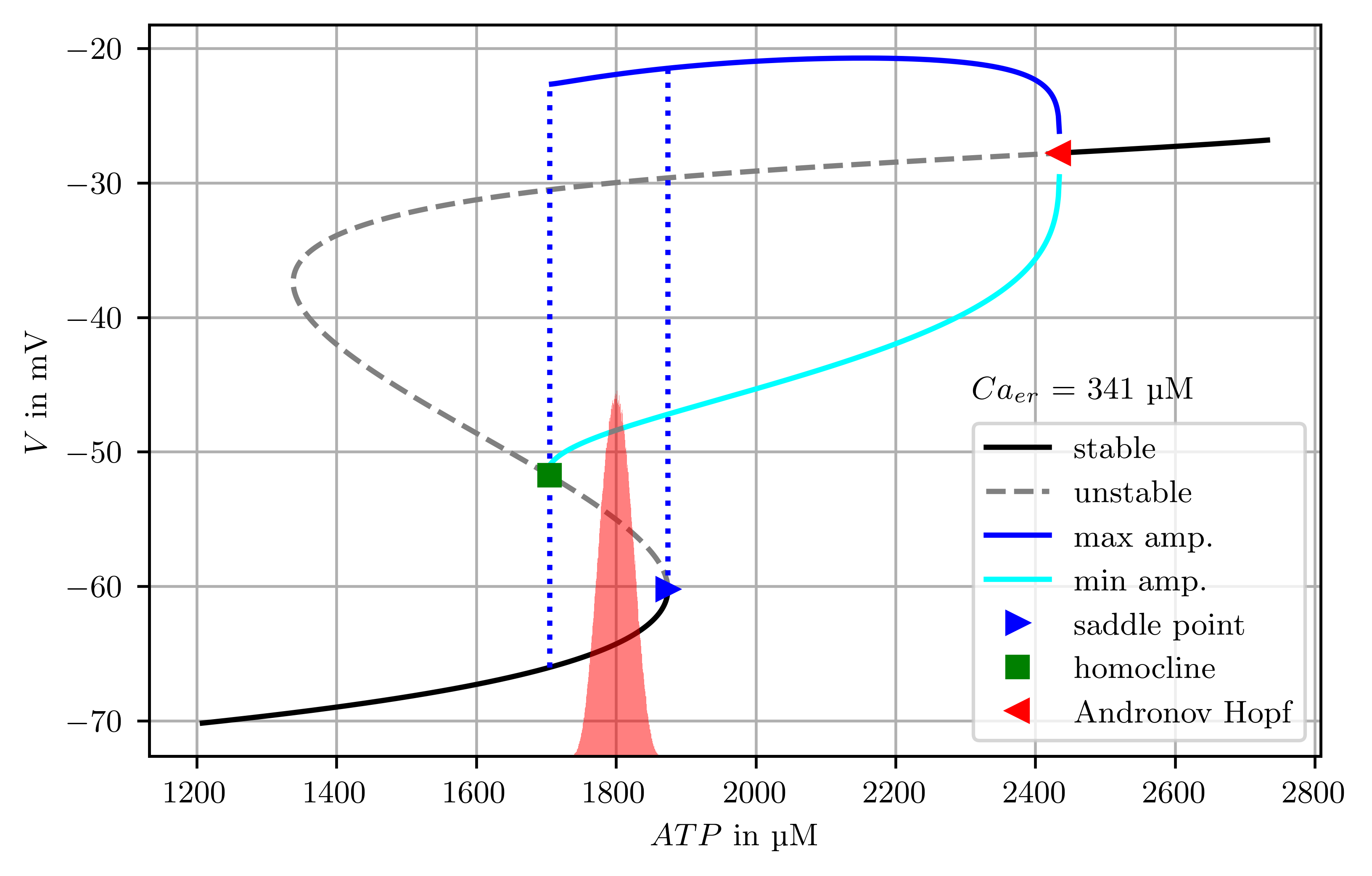}
	\caption{Bifurcation of $V$ is dependent on the concentration of $ATP$. The minimal and maximal amplitude of the LOC is indicated by dark and light blue, respectively. Sampling of Beta distribution for FgOC is indicated by the red shaded area.}
	\label{fig:cellbifurcation}
\end{figure}

As illustrated in Figure \ref{fig:cellTimeint}, a saddle homoclinic orbit and saddle node bifurcation are evident. These bifurcations are a consequence of the bifurcation parameter, $ATP$. The corresponding bifurcation diagram of $V$ over $ATP$ is illustrated in Figure \ref{fig:cellbifurcation}. 
For $ATP$ values up to $\qty{1874}{\micro M}$, a stable equilibrium is present, followed by a saddle node bifurcation. This phenomenon entails the merging of a stable and an unstable equilibrium point, which ultimately results in the cancellation of both equilibrium points and the emergence of a single unstable equilibrium point. In the region above the saddle node bifurcation until $\qty{2434}{\micro M}$, the system exhibits a single unstable equilibrium point with LCO. The maximum and minimum amplitudes are indicated by blue and cyan lines, respectively. At $ATP=\qty{2434}{\micro M}$, a supercritical Andronov-Hopf bifurcations is present, which is followed by the emergence of a stable equilibrium point. An Andronov-Hopf bifurcation occurs when a change in the bifurcation parameter causes a limit cycle to transition to a stable equilibrium point, or vice versa. A bistable regime exists for values of $ATP$ between $\qty{1704}{\micro M}$ and $\qty{1874}{\micro M}$, wherein the system is subject to a saddle homoclinic orbit bifurcation at the value of $ATP$ at $\qty{1704}{\micro M}$. In this type of bifurcation, the bifurcation parameter is decreased, resulting in an expansion of the limit cycle orbit until it reaches the homoclinic orbit. A homoclinic orbit is defined as an orbit that possesses the same equilibrium point at both its initial and final points. Subsequently, the system diverges towards a stable equilibrium point. The behavior of the full IOM model (see Fig. \ref{fig:cellTimeint}) occurs in the bistable segment, where equilibrium is reached for increasing $ATP$ values and LCO occur for decreasing values. Each LCO corresponds to a burst, as illustrated in Figure \ref{fig:cellExpSim} on the right. For more information on bifurcation, please refer to \cite{strogatzNonlinearDynamicsChaos2019,izhikevichDynamicalSystemsNeuroscience2006}.

In experimental settings, precise control of $ATP$ concentrations is not feasible, leading to inherent uncertainty in the resulting LCO. We will analyze this using the FgPC method. In this instance, we posit a normal distribution and model the uncertainty based on the assumption that the probability density function of the $ATP$ is $\text{Beta}(5,5,1730,1870)$. A total of $1e6$ samples were drawn from this input distribution. These results are presented in Figure \ref{fig:cellbifurcation}. 

It should be noted that the FgPC method is unable to capture the transition from the limit cycle to the equilibrium point. This is due to a discontinuity in the constant coefficients of the HB, which is caused by the saddle homoclinic orbit bifurcation point. For non-continuous functions, only a special gPC method exists so far, which is why FgPC reaches its limits here \cite{wanMultiElementGeneralizedPolynomial2006}. Furthermore, we observed in our experimental trails that the root-finding algorithm of the HB method of the MC method encountered difficulties in identifying a solution for samples situated in proximity to the bifurcation point. These values are subsequently employed for a comparative analysis with the FgPC, and thus we selected $\qty{1730}{\micro M}$ as the minimum ATP value.

As illustrated in \eqref{eqn:electricV} through \eqref{eqn:electricc}, the electrical system is not subjected to any external excitation.
The $ATP$ level is of significant importance because it governs the behavior of the cell membrane potential and, thus, also the electrical system in the entire IOM. As mentioned above, the ATP value is set to a fixed value because it changes only marginally during one limit cycle of the electrical system. If this value is decreased within the bistable range, the resulting limit cycle comes closer to the saddle point, resulting in a corresponding alteration in its period.
This phenomenon can be observed once more in Figure \ref{fig:cellExpSim} on the right, where the distance between the peaks of the bursts increases with time. 
Therefore, we may conclude that this is a self-excited system, which is why we have to utilize \eqref{eqn:solutionFgPC_selfexcited} for each state variable. Given that all variables are interconnected, we have one uncertain base frequency, which introduces additional $N+1$ unknown coefficients. As previously stated in Section \ref{sec:FgPC}, we set the phase information of the membrane potential $V$ to an arbitrary value. This value is calculated by performing a time integration of the electrical system with the nominal value of the $ATP$ range and extracting the phase information through a FFT. 

\begin{figure}[htbp]
	\centering
	\includegraphics{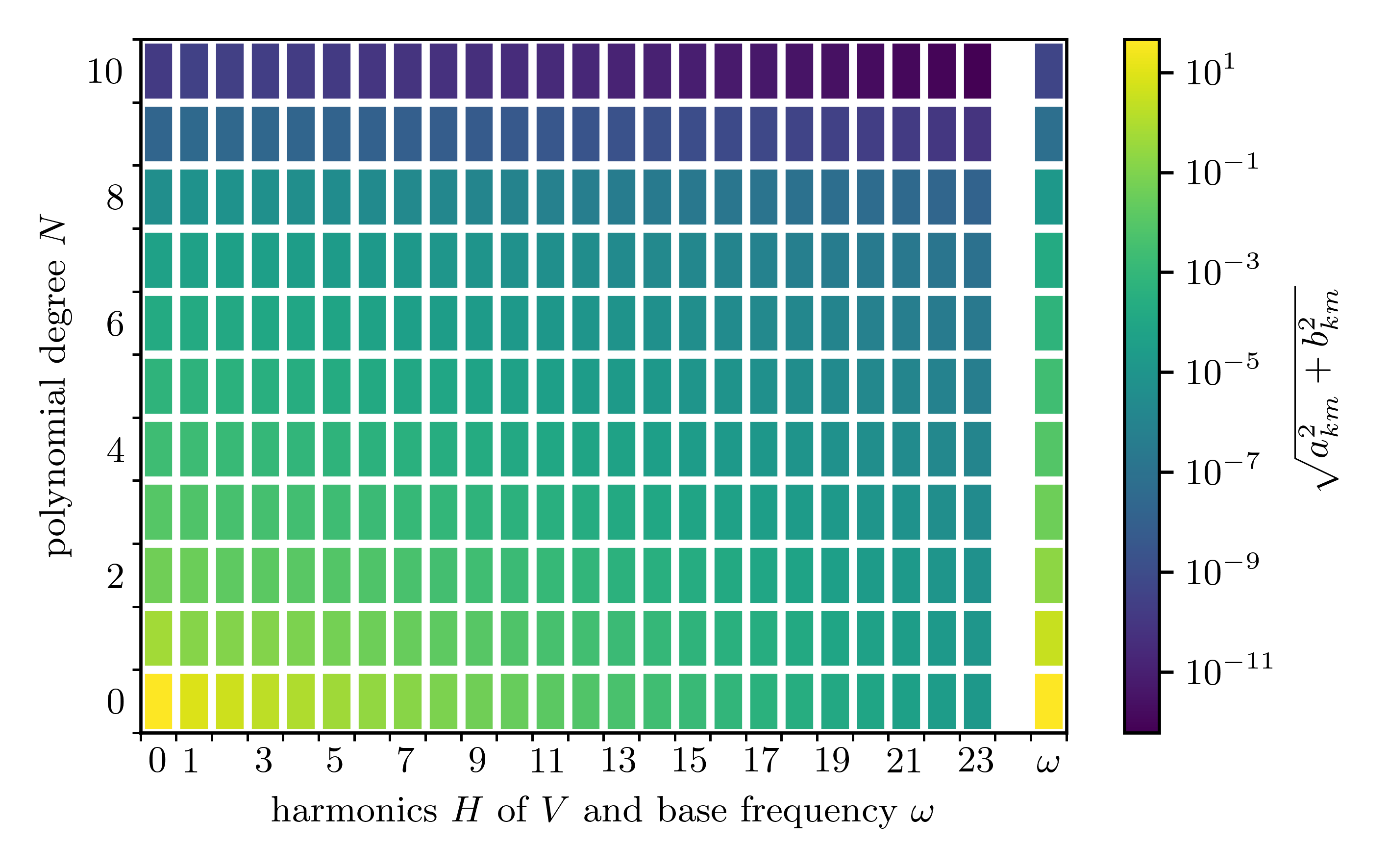}
	\caption{Magnitude of each coefficient of $V$ for every combination of the harmonic $H$ and polynomial degree $N$ and of the coefficients of the base frequency $\omega$ for each $N$ for the electrical system}
	\label{fig:cellCoeffGrid}
\end{figure}

Similar to the Duffing system, the FgPC coefficients are illustrated in \ref{fig:cellCoeffGrid} as a grid of polynomial degree over harmonics. The coefficients belong to the membrane potential $V$. Since the electrical system is self-excited, the coefficients of the base frequency are illustrated as well. Once again, the decrease in the magnitude of the coefficients is evident for increasing harmonics and polynomial degree, which indicates the existence of a sparse pattern.

\begin{figure}[htbp]
	\centering
	\includegraphics{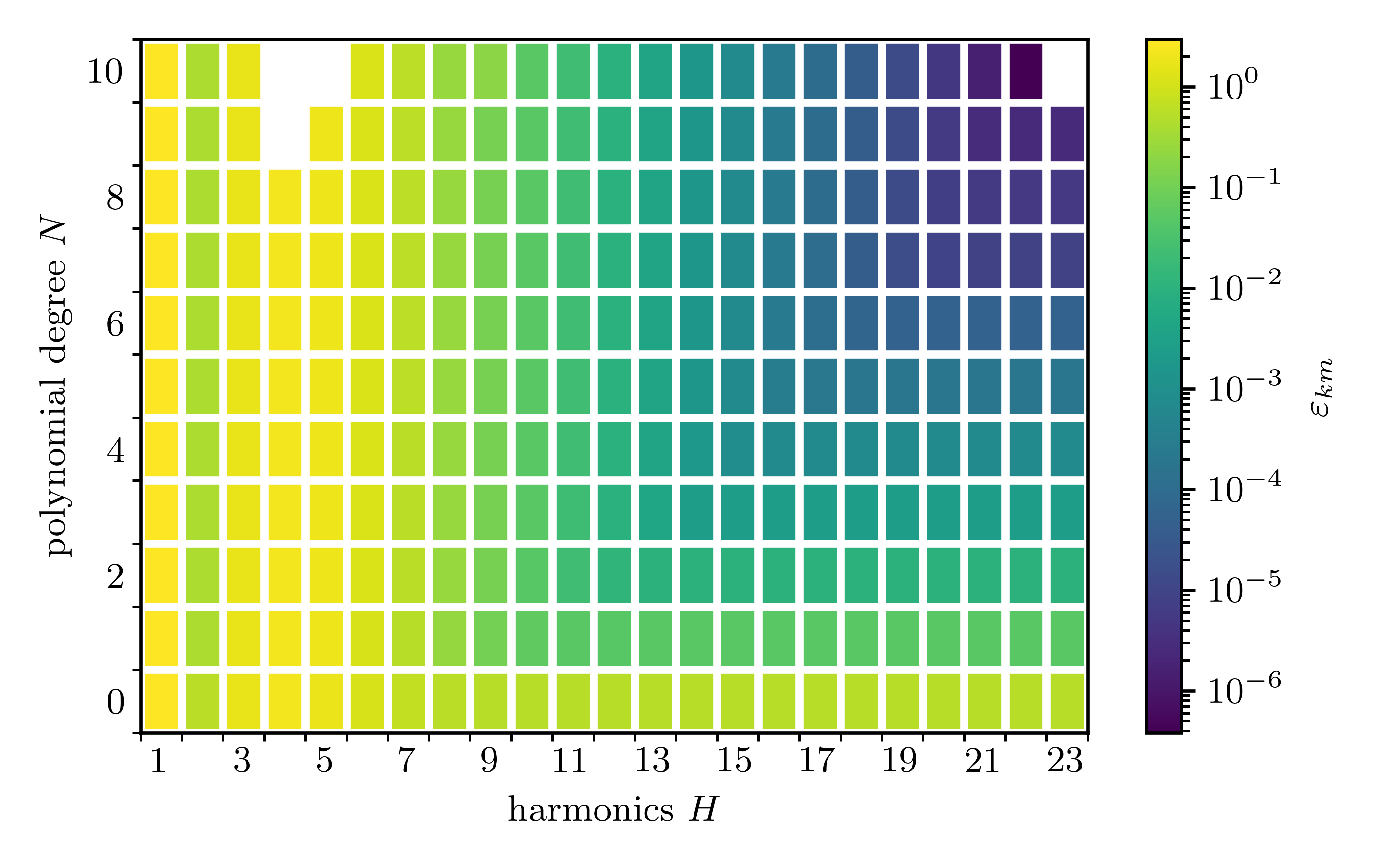}
	\caption{Total relative error of the FgPC method for every combination of harmonic $H$ and polynomial degree $N$ for the electrical system}
	\label{fig:cellErrorMap}
\end{figure}

We also conducted a convergence study for the electrical system using \eqref{eqn:relError}. As three variables are at hand, we sum up the errors for each variable and obtain the relative error map, illustrated in Figure \ref{fig:cellErrorMap}. The solution with $23$ harmonics and a polynomial degree of $10$ is used as a reference. The root-finding algorithm did not find a solution for the combination of $H=4-5$ and $N=9-10$, which we attribute to inadequate initial values, resulting in white boxes in the error map. 

\begin{figure}[htbp]
	\centering
	\includegraphics{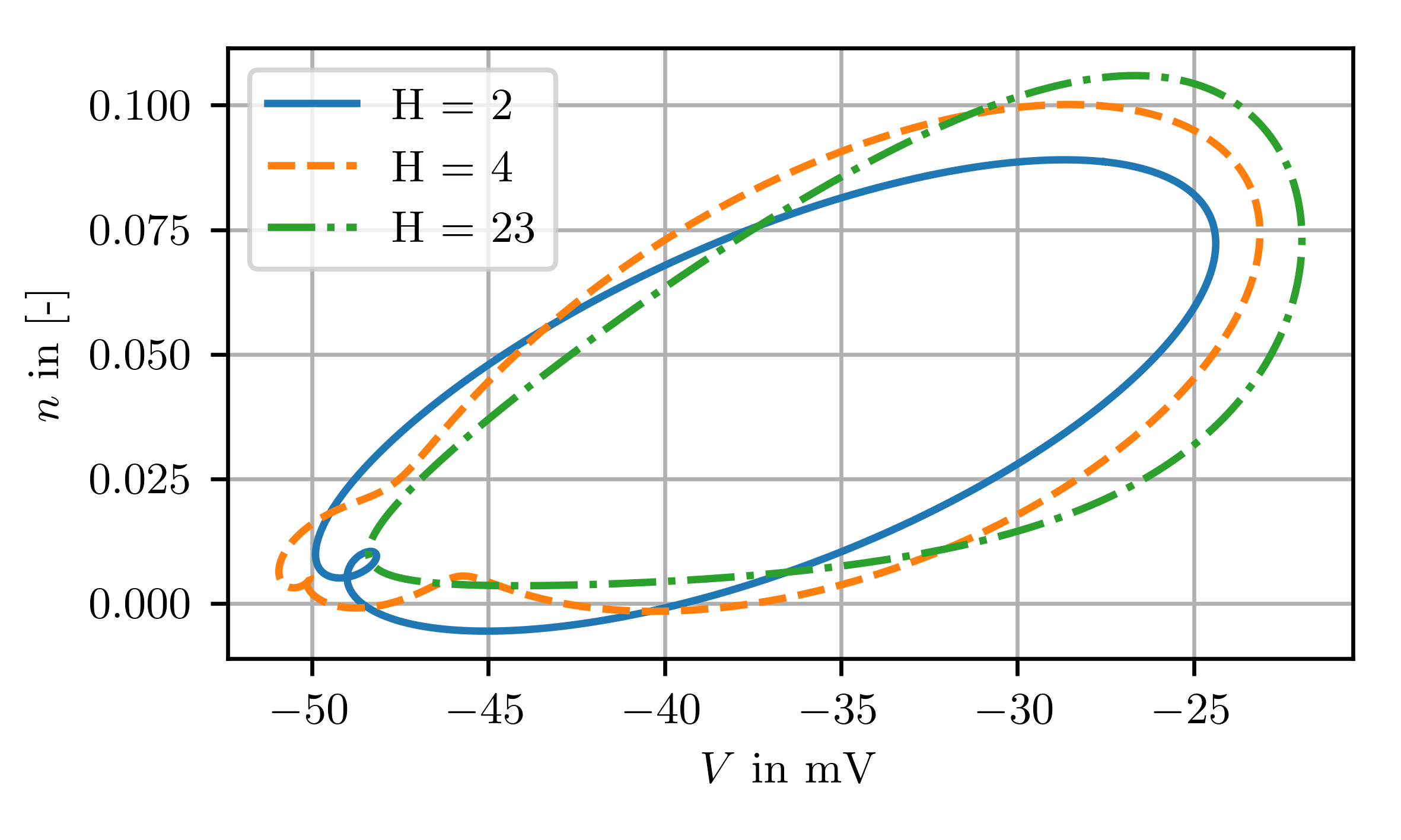}
	\caption{Phase portrait of electrical system with $n$ over $V$ for different harmonics}
	\label{fig:cellPhaseportraitCombination}
\end{figure}

As can be observed on the right-hand side, the anticipated behavior is manifest. The initial surprise lies in the observation, that the error increases after $H=2$, rather then continuing to decline as one would be anticipated. This phenomenon can be observed for harmonics greater than four. To gain further insight, we examine the behavior of different harmonics. For this purpose, we used HB with $ATP=\qty{1800}{\micro M}$. The results are presented in the phase portrait of $n$ over $V$, which is illustrated in Figure \ref{fig:cellPhaseportraitCombination}. The error \eqref{eqn:relError} is integrated over one period before comparing its result to the integral of the reference solution. With only two harmonics, the solution more closely resembles the reference solution with 23 harmonics than with four. This behavior is also captured by the error. As the number of harmonics is increased, the behavior becomes more similar to that of the true solution. 

\begin{figure}[htb]
	\centering
	\begin{subfigure}{0.48\textwidth}
		\centering
		\includegraphics{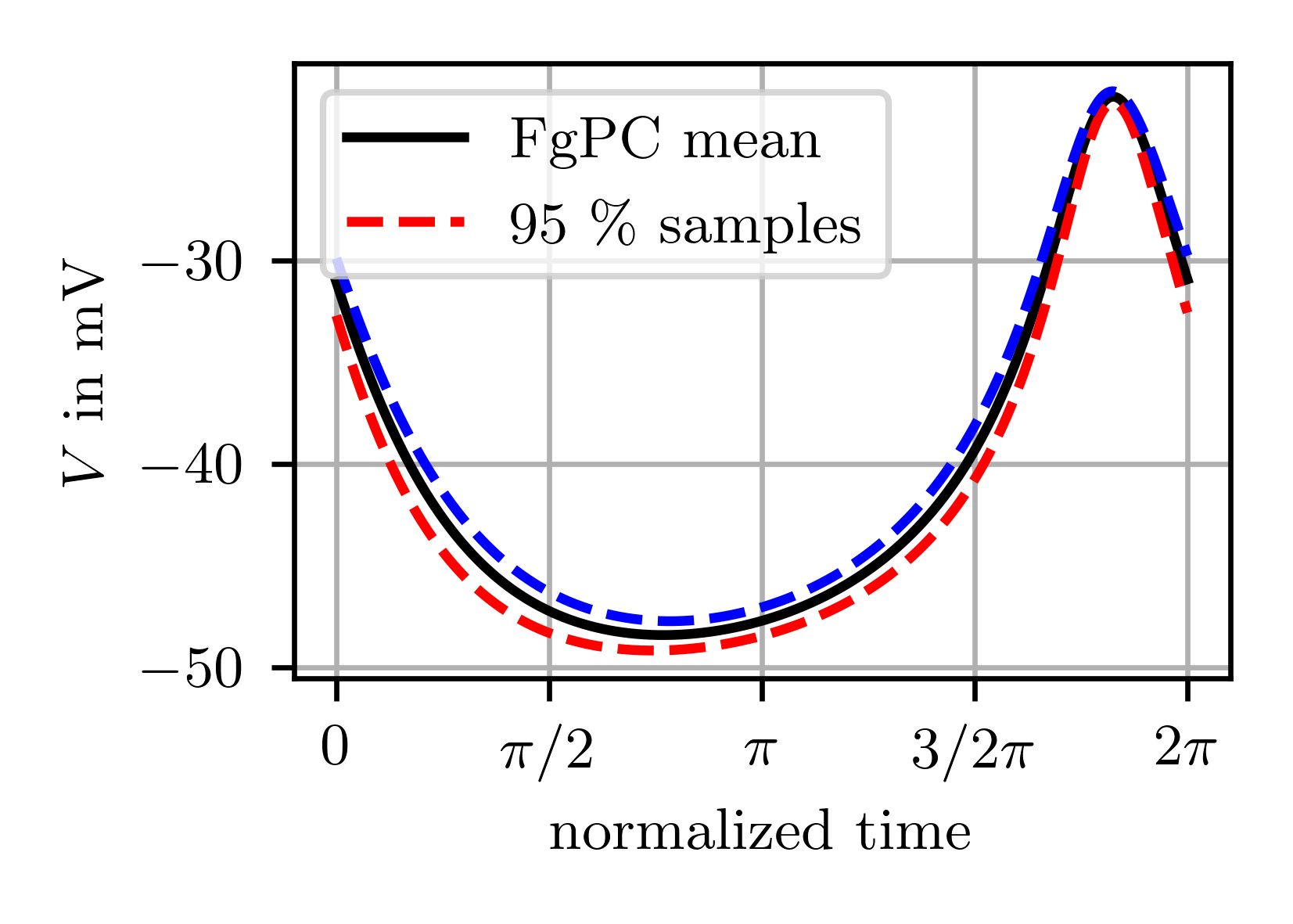}
		\caption{Mean value and interval containing 95 \% of all samples of $V$ over one period}
		\label{fig:cellPositionV}
	\end{subfigure}
	\hfill
	\begin{subfigure}{0.48\textwidth}
		\centering
		\includegraphics{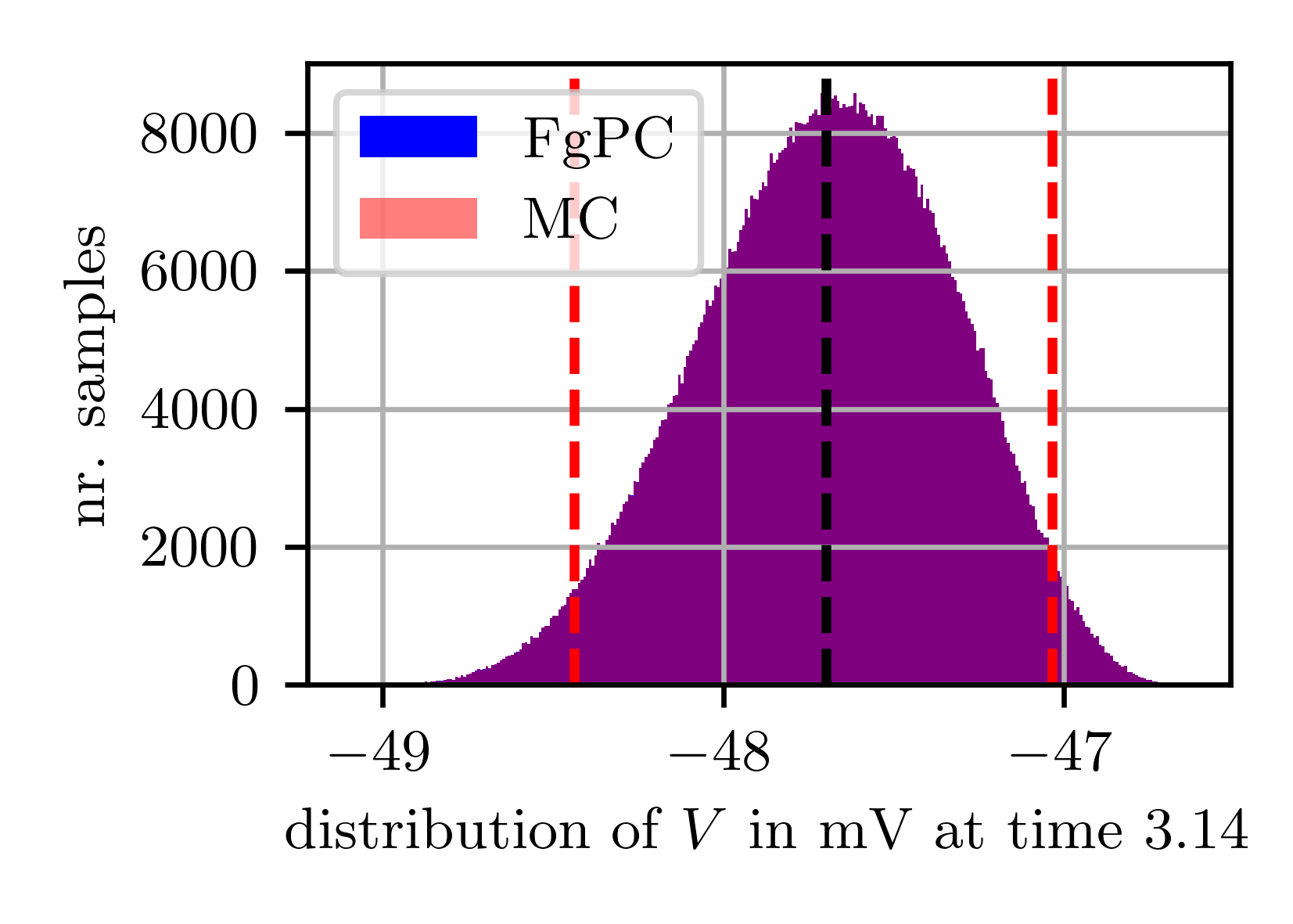}
		\caption{Distribution at normalized time $\pi$ of the FgPC and MC method, with the mean value (black dashed line) and the interval containing 95 \% of all samples (red dashed line)}
		\label{fig:cellPosVDis}
	\end{subfigure}
	\par \medskip
	\centering
	\begin{subfigure}{0.48\textwidth}
		\centering
		\includegraphics{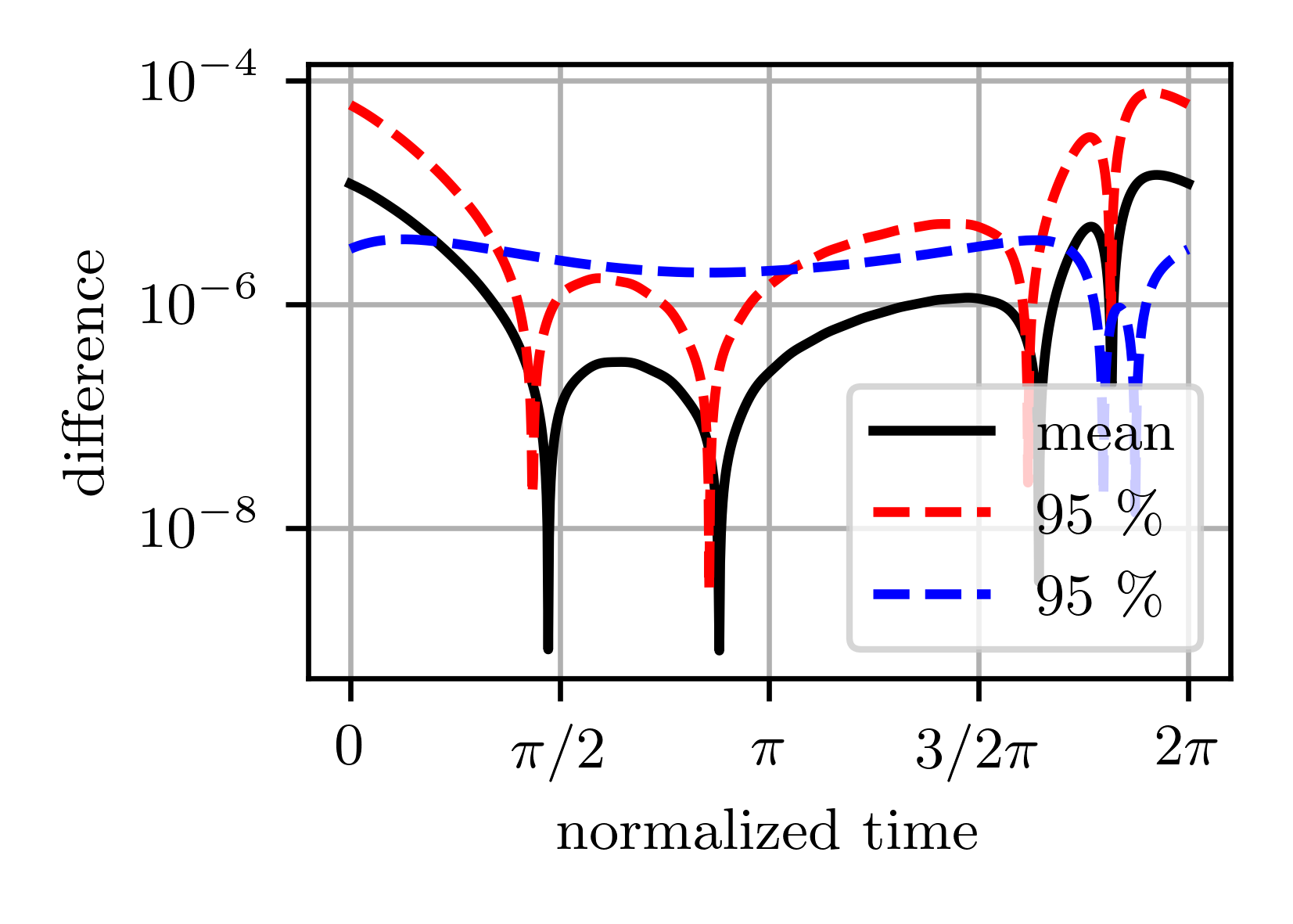}
		\caption{Difference of the FgPC and MC method for the mean value and the interval containing 95 \% of all samples of $V$ over one period}
		\label{fig:cellPosDiff}
	\end{subfigure}
	\hfill
	\begin{subfigure}{0.48\textwidth}
		\centering
		\includegraphics{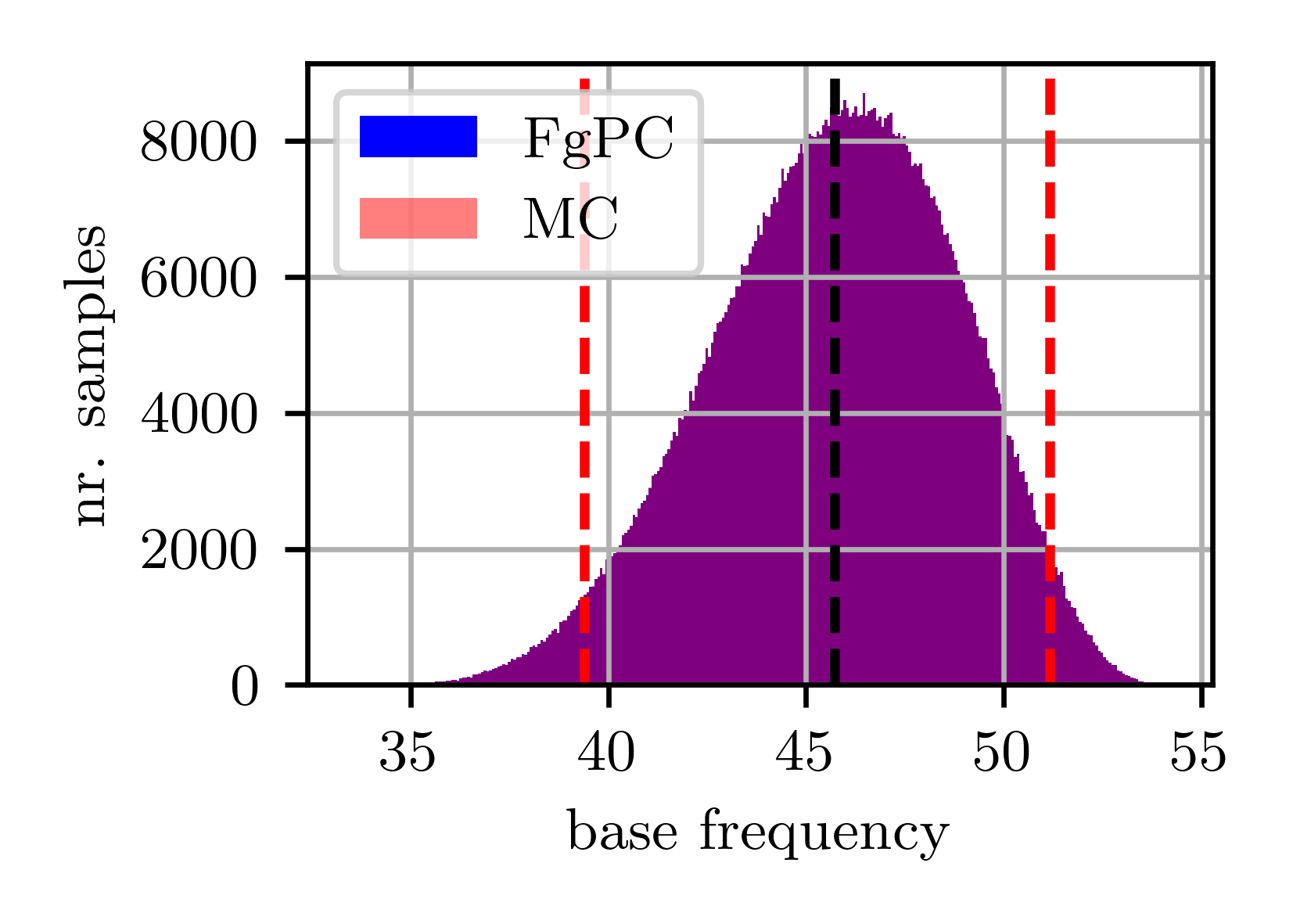}
		\caption{Distribution of the base frequency of the FgPC and MC method, with the mean value (black dashed line) and the interval containing 95 \% of all samples (red dashed line)}
		\label{fig:cellDistOmega}
	\end{subfigure}
	\caption{FgPC solution of the electrical system with $H=23$ and $N=11$}
	\label{fig:cellSystem}
\end{figure}

\begin{figure}[htb]
	\centering
	\includegraphics{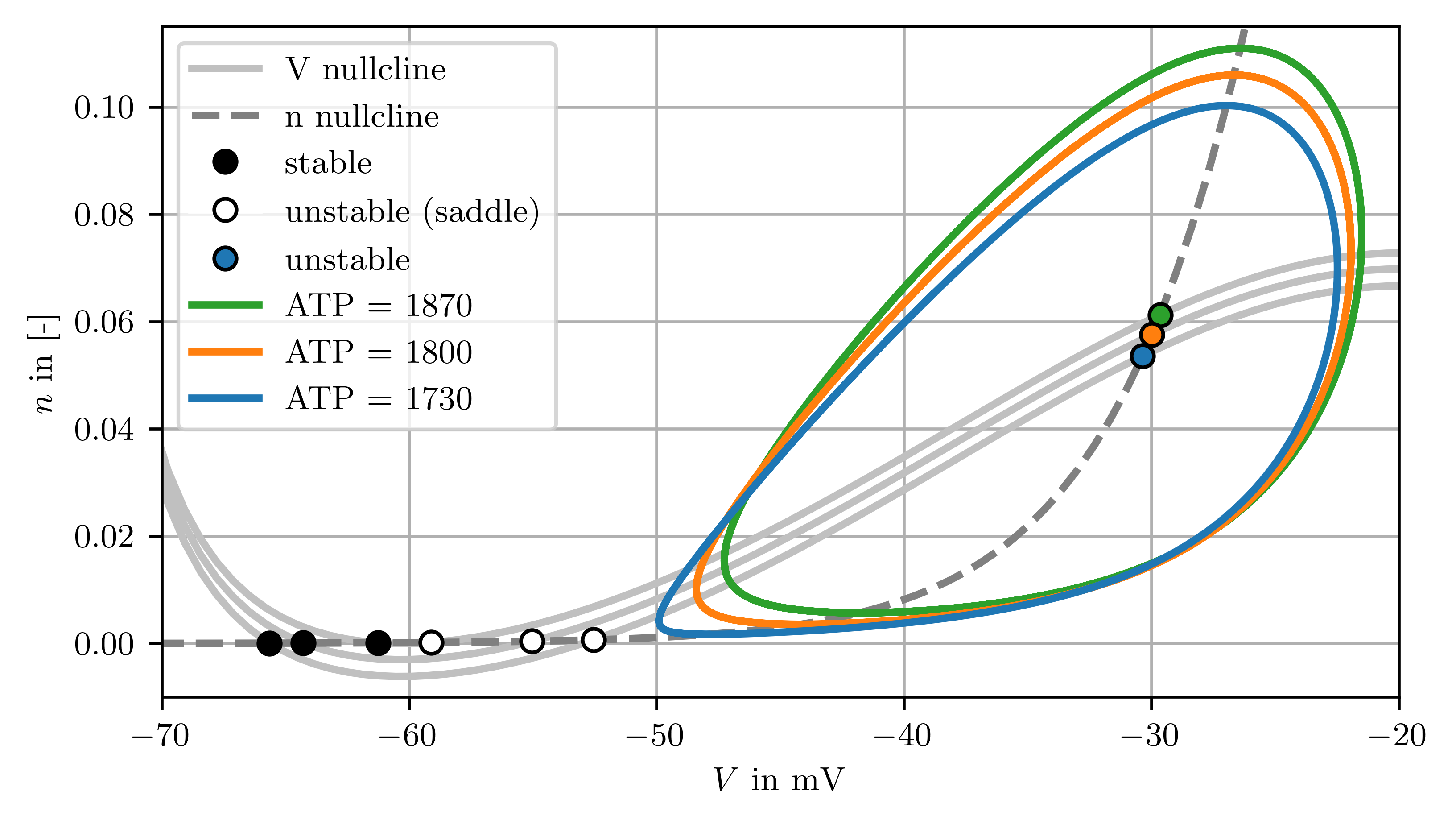}
	\caption{$n$ - $V$ plane with $V$ and $n$ nullclines of the electrical system and corresponding stable and unstable equilibrium points. Around right unstable equilibrium point limit cycles for different $ATP$ values are obtained by time integration.}
	\label{fig:cellLimitCycle}
\end{figure}

With the FgPC coefficients available for analysis, we proceed to evaluate the electrical model for all samples that have been drawn. As with the Duffing oscillator, the same stochastic quantities are employed, namely the mean value and the 95 \% sample coverage interval. In order to facilitate a presentation over the course of one oscillation period, we introduce the concept of normalized time, denoted by the range $[0,2\pi]$. This interval is then multiplied by the sample period, $T$, and the result is expressed in terms of the corresponding base frequency $\omega$. Over this normalized time interval, the stochastic quantities are evaluated at equidistant time points. For illustrative purposes, we select $V$ as an example. The behavior over one period is illustrated in Figure \ref{fig:cellPositionV}. In contrast to the Duffing oscillator, there is no crossing between the mean and the limits of the 95 \% coverage interval, since the variable does not cross zero. 
Figure \ref{fig:cellPosDiff} illustrates the discrepancy between the FgPC and MC models, which remains relatively small over the course of normalized time. Given the established compatibility of the FgPC with the MC method, it is possible to examine marginal distributions at different time points. Once more, as an illustrative example, the distribution of $V$ for a normalized time of $\pi$ is selected. 
In comparison to the input distribution of $ATP$ illustrated in Figure \ref{fig:cellbifurcation}, we observe that the sample concentration is oriented towards the right side. This shift is also evident in the distribution of the base frequency, as illustrated in Figure \ref{fig:cellDistOmega}. This shift can be attributed to the presence of a saddle point within the electrical system, which is explained in the following. Figure \ref{fig:cellLimitCycle} depicts the limit cycle for varying $ATP$ values, alongside the nullcline of $V$ and $n$ and the stable and unstable equilibrium points. The nullcline of $Ca$ is contained in that of $V$. For ATP values in proximity to the bifurcation point, the limit cycle nestles up to the saddle point, which is accompanied by a notable increase in the period duration. Here, minor alternations in $ATP$ result in a more pronounced change in the period and, consequently, the base frequency and lower $V$ and $n$ values. This expands the distribution to the left and creates a concentration on the right. 

\begin{figure}[htb]
	\centering
	\includegraphics[width =  \textwidth]{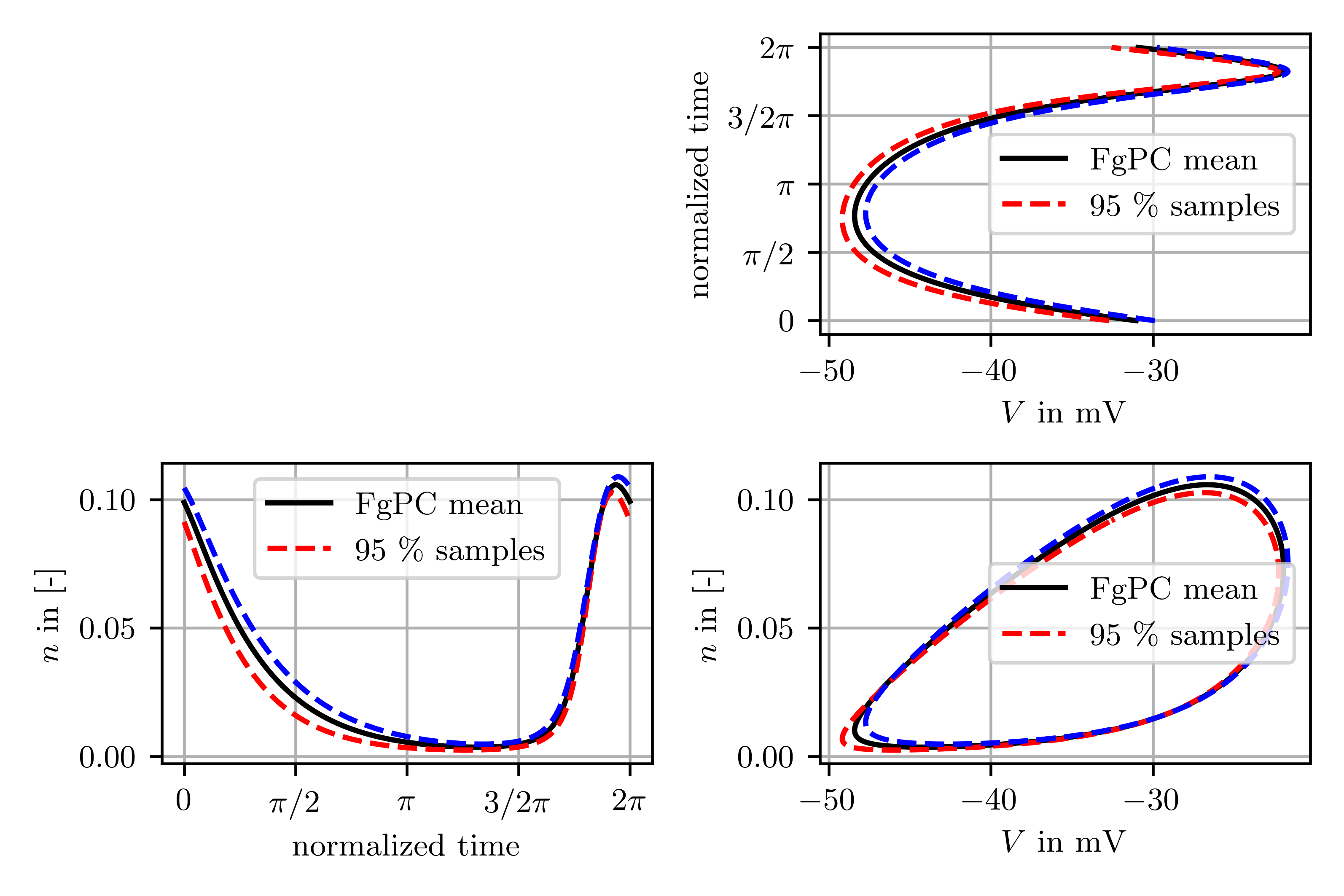}
	\caption{Phase portrait of the mean value and the interval containing 95 \% of all samples of the electrical system}
	\label{fig:cellPhaseportrait}
\end{figure}

This behavior is also captured in the LCO calculated with the FgPC. This is shown in Figure \ref{fig:cellPhaseportrait}, where the bottom left represents the behavior of $n$ over one oscillation. Similar to the Duffing oscillator, the behavior of $V$ is shown at the top right as visual support. With these two plots, it is possible to construct the resulting limit cycle at the bottom right. Here, we are able to see the same shifts as in Figure \ref{fig:cellLimitCycle}. The big difference here is that we are able to extract the marginal distributions for given time points. The distribution of the base frequency is an additional benefit of the FgPC method.

The FgPC method served as a foundation for efficient uncertainty quantification of the electrophysiological properties of $\beta$ cells. A comprehensive interpretation, encompassing the biological functionality of the cell, is the subject of future research.

\section{Summary and Discussion}
\label{sec:summary}

This paper focuses on stochastic predictions of limit cycle oscillations of nonlinear systems under uncertainties. For this purpose, we used the FgPC method and integrated it with a deflation technique to identify multiple solutions of the specified system parameter set.

We applied the FgPC method to the Duffing oscillator to introduce the stochastic quantities used throughout this study - and to facilitate an understanding of their behavior. The FgPC method yielded results that were comparable to those obtained by a Monte Carlo method with a classical harmonic balance approach. The computational time was significantly reduced, and the discrepancy between the two methods was minimal. 

To illustrate the FgPC method's suitability for more intricate and pertinent problems, we applied it to the electrophysiological model of a pancreatic $\beta$ cell. This allowed us to anticipate the stochasticity of limit cycle oscillations and the base frequency. This outcome can be utilized to enhance the modeling of experiments. Additionally, with this illustration, we also demonstrated that the method can effectively handle systems with a higher harmonic and polynomial degree.

Looking at the coefficients in both cases, a sparsity pattern becomes evident. This is an important finding - leading to the conclusion that the FgPC method can exploit this sparsity for an adaptive algorithm. This can improve the efficiency of the root-finding algorithm by reducing the number of unknowns to search. This idea will be investigated in future work.

Given the applicability of the FgPC method to a wide range of non-linear problems, we intend to apply it to other engineering-relevant problems, such as FEM models with nonlinearities and uncertain model parameters. Additionally, we also intend to explore the possibilities of the stochastic interpretation of single and multiple solutions that appear with parameter changes, as evidenced by the parameter set of the Duffing oscillator.

\section*{CRediT authorship contribution statement}
\textbf{Lars de Jong:} Conceptualization, Methodology, Software, Validation, Visualization, Writing – original draft, Writing – review and editing
\textbf{Paula Clasen:} Software, Validation, Visualization, Writing – review and editing
\textbf{Michael Müller:} Conceptualization, Methodology, Supervision, Writing – review and editing
\textbf{Ulrich Römer:} Conceptualization, Methodology, Supervision, Writing – review and editing

\section*{Declaration of competing interest} 
The authors declare that they have no known competing financial interests or personal relationships that could have appeared to influence the work reported in this paper. 

\section*{Code availability}
Code is available under \cite{de_jong_2024_13359870}. 

\section*{Funding}
This research did not receive any specific grant from funding agencies in the public, commercial, or not-for-profit sectors.


\bibliographystyle{elsarticle-num-names} 
\bibliography{paper}





\end{document}